\documentclass[letterpaper,twocolumn,10pt]{article}
\usepackage{usenix-2020-09}

\usepackage{hyperref}
\usepackage[numbers]{natbib}
\usepackage{microtype}
\usepackage{amsthm}
\usepackage{amssymb}
\usepackage{graphicx}
\usepackage{subcaption}
\usepackage{wrapfig}
\usepackage{booktabs}
\usepackage{url}

\usepackage{amsmath}
\usepackage{mathtools}
\usepackage{amsfonts,bm}

\usepackage{xcolor}
\usepackage[normalem]{ulem}
\usepackage{enumitem}
\usepackage{newfloat}
\usepackage[most]{tcolorbox}
\usepackage{algorithm}
\usepackage{algpseudocode}

\usepackage[capitalize,noabbrev]{cleveref}
\makeatletter
\let\old@ref\ref
\makeatother
\AtBeginDocument{
\renewcommand{\ref}[1]{%
  \PackageError{main}{Use {\string\cref} instead of {\string\ref}!}{Please always use {\string\cref} for consistency.}%
  {\textbf{\color{red} DON'T USE REF; USE CREF}}
}
}

\usepackage{xparse}
\usepackage{xfp}
\usepackage{xspace}

\usepackage{booktabs, multirow, float} %

\makeatletter
\renewcommand\paragraph{\@startsection{paragraph}{4}{\z@}{1ex}{-1em}{\normalfont\normalsize\bfseries}}
\makeatother

\newcommand{\nsattack}{Netflix Prize attack\xspace}

\newcommand{\matchfrac}{\pi}

\newcommand{\tpr}{\text{TPR}}
\newcommand{\fpir}{\text{FPIR}}
\newcommand{\fmr}{\text{FMR}}
\DeclareMathOperator{\precision}{\text{Precision}}

\NewDocumentCommand{\metric}{o o}{%
  \IfNoValueTF{#1}%
    {Recall@Precision\xspace}%
    {\IfNoValueTF{#2}%
      {Recall@\fpeval{#1 * 100}\% Precision\xspace}%
      {Recall@\fpeval{#1 * 100}\% Precision ($\matchfrac = #2$)}\xspace%
    }%
}

\theoremstyle{plain}

\theoremstyle{definition}

\theoremstyle{remark}

\newcommand{\code}[1]{\texttt{#1}}

\newif\ifshowtopics
\showtopicsfalse  %

\newif\ifshowcomments
\showcommentstrue  %
\ifshowcomments
  \newcommand{\daniel}[1]{\textcolor{blue}{\{daniel: #1\}}}
  \newcommand{\dpaleka}[1]{\textcolor{blue}{\{daniel: #1\}}}
  \newcommand{\simon}[1]{\textcolor{orange}{\{simon: #1\}}}
  \newcommand{\florian}[1]{\textcolor{purple}{\{florian: #1\}}}
  \newcommand{\joshua}[1]{\textcolor{teal}{\{joshua: #1\}}}
  \newcommand{\michael}[1]{\textcolor{olive}{\{michael: #1\}}}
  \newcommand{\nc}[1]{\textcolor{magenta}{\{nc: #1\}}}
  \newcommand{\edit}[2]{#2}
\else
  \newcommand{\daniel}[1]{}
  \newcommand{\dpaleka}[1]{}
  \newcommand{\simon}[1]{}
  \newcommand{\florian}[1]{}
  \newcommand{\joshua}[1]{}
  \newcommand{\michael}[1]{}
  \newcommand{\nc}[1]{}
  \newcommand{\edit}[2]{#2}
\fi

\definecolor{systemdark}{HTML}{7697AD}
\definecolor{systembright}{HTML}{FFFFFF}
\definecolor{assistantdark}{HTML}{75AC9D}
\definecolor{assistantbright}{HTML}{FFFFFF}
\definecolor{userdark}{HTML}{7697AD}
\definecolor{userbright}{HTML}{FFFFFF}

\newtcolorbox{systembox}{
    flush left,
    width = 1.0\textwidth,
    colback = systembright,
    colframe = systemdark,
    enhanced,
    fuzzy shadow = {0pt}{-2pt}{-0.5pt}{0.5pt}{black!35},
    fonttitle=\footnotesize\bfseries,
    halign title=flush left,
    title = System,
    before upper={\footnotesize},
    fontupper=\footnotesize
}
\newtcolorbox{assistantbox}{
    flush right,
    width = 1.0\textwidth,
    colback = assistantbright,
    colframe = assistantdark,
    enhanced,
    fuzzy shadow = {0pt}{-2pt}{-0.5pt}{0.5pt}{black!35},
    fonttitle=\footnotesize\bfseries,
    halign title=flush right,
    title = Assistant,
    before upper={\footnotesize},
    fontupper=\footnotesize
}
\newtcolorbox{userbox}{
    flush left,
    width = 1.0\textwidth,
    colback = userbright,
    colframe = userdark,
    enhanced,
    fuzzy shadow = {0pt}{-2pt}{-0.5pt}{0.5pt}{black!35},
    fonttitle=\footnotesize\bfseries,
    halign title=flush left,
    title = User,
    before upper={\footnotesize},
    fontupper=\footnotesize
}

\date{}

\title{\Large \bf Large-scale online deanonymization with LLMs}

\author{
\begin{tabular}[t]{c}
\begin{tabular}[t]{c@{\hspace{3em}}c}
\begin{tabular}[t]{c}{\rm Simon Lermen\textsuperscript{*}}\\ MATS\end{tabular} &
\begin{tabular}[t]{c}{\rm Daniel Paleka\textsuperscript{*}}\\ ETH Zurich\end{tabular}
\end{tabular} \\[1.5em]
\begin{tabular}[t]{c@{\hspace{2em}}c@{\hspace{2em}}c@{\hspace{2em}}c}
\begin{tabular}[t]{c}{\rm Joshua Swanson}\\ ETH Zurich\end{tabular} &
\begin{tabular}[t]{c}{\rm Michael Aerni}\\ ETH Zurich\end{tabular} &
\begin{tabular}[t]{c}{\rm Nicholas Carlini}\\ Anthropic\end{tabular} &
\begin{tabular}[t]{c}{\rm Florian Tram\`er}\\ ETH Zurich\end{tabular}
\end{tabular}
\end{tabular}
}

\begin{document}

\newlength{\figcontentwidth}
\setlength{\figcontentwidth}{\textwidth}
\newlength{\figgutterwidth}
\setlength{\figgutterwidth}{10pt}
\newlength{\figcolwidth}
\setlength{\figcolwidth}{0.083333\figcontentwidth-0.916667\figgutterwidth}

\newlength{\figtwelvecol}
\setlength{\figtwelvecol}{12\figcolwidth+11\figgutterwidth}
\newlength{\figninecolwidth}
\setlength{\figninecolwidth}{9\figcolwidth+8\figgutterwidth}
\newlength{\figsixcol}
\setlength{\figsixcol}{6\figcolwidth+5\figgutterwidth}
\newlength{\figfivecol}
\setlength{\figfivecol}{5\figcolwidth+4\figgutterwidth}
\newlength{\figfourcol}
\setlength{\figfourcol}{4\figcolwidth+3\figgutterwidth}
\newlength{\figthreecol}
\setlength{\figthreecol}{3\figcolwidth+2\figgutterwidth}
\newlength{\figonecol}
\setlength{\figonecol}{\figcolwidth}

\newlength{\figfull}
\setlength{\figfull}{\figtwelvecol}
\newlength{\fighalf}
\setlength{\fighalf}{\figsixcol}
\newlength{\figthird}
\setlength{\figthird}{\figfourcol}

\maketitle

\renewcommand{\thefootnote}{\fnsymbol{footnote}}
\footnotetext[1]{Equal contribution.}
\renewcommand{\thefootnote}{\arabic{footnote}}

\begin{abstract}
    We show that large language models can be used to perform at-scale deanonymization.
    With full Internet access, our agent
    can re-identify Hacker News users and Anthropic Interviewer participants at high precision,
    given pseudonymous online profiles and conversations alone, 
    matching what would take hours for a dedicated human investigator. 
    We then design attacks for the closed-world setting.
    Given two databases of pseudonymous individuals, each containing unstructured text written by or about that individual, we implement a scalable attack pipeline that uses LLMs to: (1) extract identity-relevant features, (2) search for candidate matches via semantic embeddings, and (3) reason over top candidates to verify matches and reduce false positives.
    Compared to \edit{prior}{classical} deanonymization work (e.g., on the Netflix prize) that required structured data \edit{or manual feature engineering}{}, our approach works directly on raw user content across arbitrary platforms.
    We construct three datasets with known ground-truth data to evaluate our attacks.
    The first links Hacker News to LinkedIn profiles, using cross-platform references that appear in the profiles.
    Our second dataset matches users across Reddit movie discussion communities;
    and the third splits a single user's Reddit history in time to create two pseudonymous profiles to be matched.
    In each setting, LLM-based methods substantially outperform classical baselines,
   achieving up to 68\% recall at 90\% precision compared to near 0\% for the best non-LLM method.
    Our results show that the practical obscurity protecting pseudonymous users online no longer holds and that threat models for online privacy need to be reconsidered.
\end{abstract}

\section{Introduction}
For decades, it has been known that individuals can be uniquely identified from surprisingly few attributes.
Sweeney's seminal work demonstrated that 87\% of the U.S. population could be uniquely identified by just zip code, birth date, and gender~\cite{sweeney2002k}.
Narayanan and Shmatikov showed that anonymous Netflix ratings could be linked to public IMDb profiles using only a handful of movie preferences~\cite{narayanan2008deanonymization}, while \citet{de2013unique} proved that four spatiotemporal points are enough to uniquely identify 95\% of individuals in mobile phone datasets.
Despite these attacks, pseudonymous online accounts (Reddit throwaways, anonymous forums, review profiles, etc) have remained largely unaffected by deanonymization attempts.
The reason is simple: applying such attacks in practice has required structured data amenable to algorithmic matching or substantial manual effort by skilled investigators reserved for high-value targets~\citep{garcia2017evidentiary}.
    
\edit{}{Deanonymization is a two-step process at heart, involving profiling an anonymous person from their posts, and then matching them to a known identity.}
\edit{}{It's well-known that large language models (LLMs) can infer personal attributes from text on online forums \cite{staab2024beyond,du2025automated, xin2025false}.}
\edit{}{Given this, it makes sense to ask: how good are LLMs at full end-to-end deanonymization, and is this a practical threat to pseudonymous accounts?}

\paragraph{Our contributions.}
We demonstrate that LLMs fundamentally change \edit{this calculus}{the picture}, enabling fully automated deanonymization attacks that operate on unstructured text at scale.
\edit{Where previous approaches required predefined feature schemas, careful data alignment, and manual verification, LLMs can}{We show this by phrasing deanonymization as a matching problem and showing LLMs can perform all steps needed to match accounts:}
extract identity-relevant signals from arbitrary \edit{prose}{text}, efficiently search over millions of candidate profiles, and reason about whether two accounts belong to the same person.
We show that the practical obscurity that has long protected pseudonymous users (the assumption that deanonymization, while theoretically possible, is too costly to execute broadly) no longer holds.

\begin{figure*}[t]
    \centering
    \includegraphics[width=\textwidth, trim={0 1cm 0 0},clip]{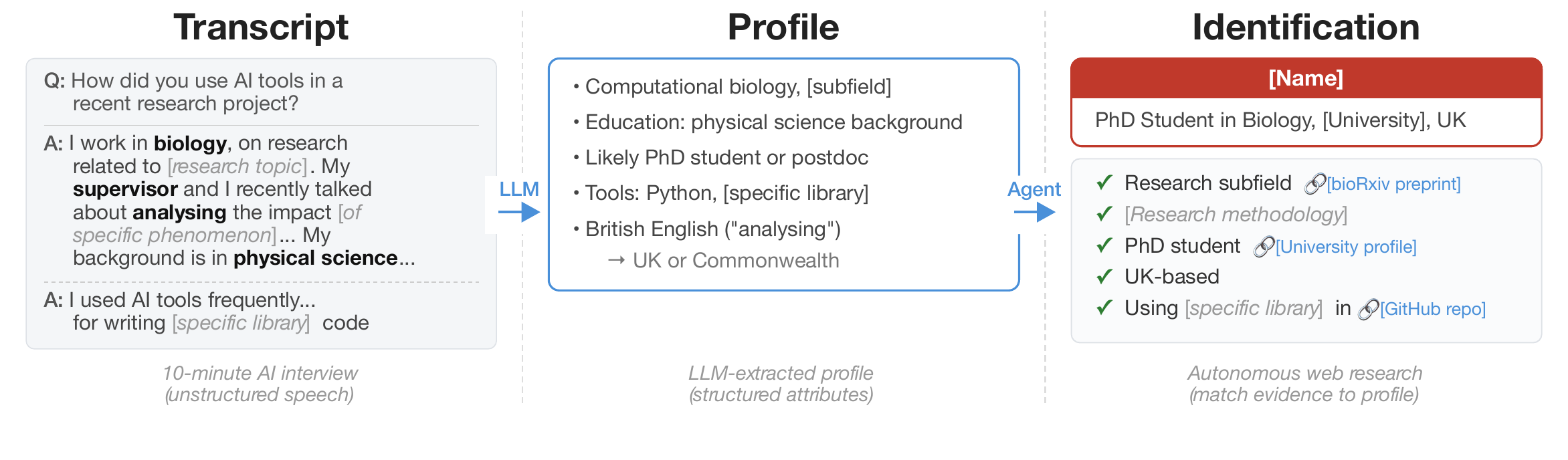}
    \caption{End-to-end deanonymization from a single interview transcript from~\cite{anthropic2025interviewer} (\emph{details altered to protect the subject's identity}). An LLM agent extracts structured identity signals from a conversation,
    autonomously searches the web to identify a candidate individual, and verifies the candidate matches all extracted claims.}
    \label{fig:anthropic_scientists_example}
\end{figure*}

We validate this thesis in three deanonymization settings%
: (1) \edit{deanonymizing}{matching} an online account to its real identity; (2) linking an identity to an unknown pseudonymous account; and (3) linking pseudonymous accounts of the same person across different platforms or time periods.
These settings capture distinct threat models (e.g., doxxing of an online account, a stalker targeting a victim, or an adversary consolidating a user's activity) and pose different technical challenges.

In the first setting, we demonstrate that state-of-the-art LLM agents already possess reasonable capabilities to perform end-to-end deanonymization fully autonomously on the open web.
This is the most challenging setup we consider: 
Given only an anonymous online profile, our LLM agents autonomously search the web, query databases, enumerate potential candidate identities, and reason over evidence to identify who the profile belongs to.
In a study of Hacker News and Reddit profiles, these agents achieve 25--67\% recall with 70--90\% precision, replicating in minutes what could take hours for a dedicated human investigator.
\Cref{fig:anthropic_scientists_example} illustrates a successful deanonymization  of a participant in the Anthropic Interviewer Dataset~\cite{anthropic2025interviewer, li2026agentic}, a collection of discussions with 125 scientists about how they integrate AI into their work. We estimate that our agents correctly re-identify at least 9/125 participants from this dataset.
\emph{This capability comes for free}: we simply prompt frontier agents with a summary of an online profile and ask the agent to uncover the identity behind it. 

To enable a more granular study of how LLMs can improve upon prior deanonymization attacks in the second and third settings (linking pseudonymous accounts to an identity or across platforms), we then design a scalable deanonymization pipeline in four LLM-augmented stages:
\begin{itemize}
\item \emph{Extract}: \edit{}{as in prior work \citep{staab2024beyond,du2025automated}}, we ask LLMs to identify and structure relevant features from unstructured posts: demographics, writing style, interests, incidental disclosures, etc.
\item \emph{Search}: we encode extracted features into dense embeddings enabling efficient search over thousands or millions of candidate profiles.
\item \emph{Reason}: we use extended reasoning on top candidates from the search step to identify the most likely match given all available context;
\item \emph{Calibrate}: we prompt LLMs to provide confidences in identified matches (either absolute or relative to other matches), which lets us calibrate the attack to a desired false positive rate.
\end{itemize}

Our pipeline substantially outperforms adaptations of classical deanonymization techniques~\cite{narayanan2008deanonymization}.
For example, we improve recall from 0.1\% to 45.1\% at 99\% precision when linking Hacker News accounts to LinkedIn profiles; and we improve recall from 0\% to 2.8\% at the same precision threshold on a challenging task of linking Reddit users across movie subreddits.
Ablations confirm that each pipeline stage contributes: in particular, the \emph{Reason} step improves recall at 99\% precision from 4.4\% (Search only) to 45.1\%.

Our work introduces an evaluation framework for large-scale deanonymization attacks, which we believe can be broadly useful for follow-up studies. 
Indeed, evaluating deanonymization attacks at scale poses inherent challenges, since ground-truth labels are difficult to obtain without compromising the privacy of real users.
As a result, previous work evaluated scale through synthetic data~\cite{narayanan2009deanonymizing} and relied on manual verification or guess-work to validate attacks on real data~\cite{narayanan2008deanonymization, cohen2022attacks}.
We propose two approaches that balance real-world validity with research ethics.
Our first approach identifies profiles that are \emph{not} anonymous (for instance, a Hacker News account whose ``about'' field links to a LinkedIn profile) and then renders them pseudonymous for evaluation purposes by removing all direct identifiers. We then measure whether our attacks can recover the removed link.
Although this approach may not capture the behavior of the most privacy-conscious users, it provides verifiable ground truth at scale and enables rigorous comparison across methods.
Our second approach splits a single user's activity across communities or across time, synthetically creating two user profiles with a known link which our attacks then try to infer. This mitigates selection bias concerns but comes with other limitations discussed later.

\paragraph{Implications.}
Our findings have significant implications for online privacy.
The average online user has long operated under an implicit threat model where they have assumed pseudonymity provides adequate protection because targeted deanonymization would require extensive effort.
LLMs invalidate this assumption.
\edit{They do not necessarily do so by exceeding human \emph{capability}---the signals our models exploit are the same signals that a skilled investigator would recognize---but by reducing \emph{cost}.}{As also observed by \citet{staab2024beyond}, they do not necessarily do so by exceeding human \emph{capability}---the signals our models exploit are the same signals that a skilled investigator would recognize---but by reducing \emph{cost}. Our experiments provide quantitative evidence for this in the deanonymization setting.}
\edit{}{We discuss our relationship to related work on online profiling and privacy in \cref{sec:related_work}; and address memorization concerns in \cref{sec:discussion}.}

Past research identified that generally the only effective mitigation against privacy attacks applied to ``anonymized'' structured data was simply not to release such data at all~\cite{narayanan2008deanonymization}. Our findings suggest that unstructured text data deserves similar reconsideration. However, this text is the very content that makes online communities valuable. We therefore argue that privacy expectations, platform policies, and social norms that govern pseudonymous participation online require reconsideration.
We hope that this work can spark that conversation.

\section{LLM agents can autonomously re-identify anonymous online profiles}
\label{sec:study_agents}

To best highlight the paradigm shift that LLMs bring in online deanonymization, we begin our study by directly evaluating whether frontier LLMs can autonomously perform end-to-end deanonymization---from anonymous profile to real-world identity.
\edit{This is a setting that has so far only been amenable to manual attacks, and included manually collecting hints and clues from online posts and performing strategic searches.}{This setting has so far required a significant manual component even if LLMs are used for attribute extraction~\citep{du2025automated}.}

In order to evaluate whether LLMs can carry out this capability, we first need to build an evaluation set of online profiles along with a known ground truth identity.
We do this by collecting \emph{non-anonymous} online profiles and then \edit{explicitly removing}{using an LLM to remove} any directly identifying information.
\Cref{tab:anonymization} summarizes the anonymization rules we use: we remove personal URLs, social media handles, GitHub repositories (which directly identify the owner) and retain institutions, demographics, interests, and names of connected people (which are too broad to identify the profile owner alone).
The goal is to create a task that would be nontrivial for a motivated \edit{human with web search access}{human}. We then test whether LLM agents can solve this task by performing web searches.
This artificial anonymization is only needed for evaluation (to establish ground truth); any real attack would target profiles that are already pseudonymous.

We provide the agent with a text description of a person (derived from their pseudonymous posts) and ask it to determine their real identity by searching the web, cross-referencing sources, and reasoning over evidence. 

Our pipeline proceeds as follows:
\begin{enumerate}
    \item Summarize the user's posts into a profile containing both stated and inferred facts about demographics, career history, interests, and so on. %
    \item Generate a search prompt from the profile.
    Anonymize the search prompt by removing names, handles, and unique identifiers that would allow for a direct search.
    \item Pass the anonymized search prompt to an LLM agent with web search tools. 
    The agent autonomously searches, reasons over findings, and attempts to identify the user.
\end{enumerate}
Following \citet{li2026agentic}, to prevent misuse, we describe our attack at a high level, and do not publish the agent, exact prompts, or tool configurations used. Running the agent on each profile costs us between \$1--\$4; making the cost of the experiments below less than \$2000.

\subsection{Results}

We first evaluate our agentic pipeline on three ground-truth datasets based on Hacker News and Reddit profiles. 
In the interest of research ethics, \textbf{we do not evaluate our method on any truly pseudonymous accounts on Hacker News and Reddit}.
We additionally report on the deanonymization of genuinely pseudonymous Anthropic Interviewer transcripts that were the subject of prior deanonymization research~\citep{li2026agentic}.

\paragraph{Hacker News $\to$ LinkedIn.}
We collect 338 Hacker News (HN) users who linked a LinkedIn profile in their publicly-visible HN bio, providing verified real-world identities as ground truth.
We first summarize each user's HN activity (comments and stories)
into a structured profile. Then we create a search prompt and anonymize it (see \Cref{app:anonymization} for details), 
and pass it to the agent.
The agent correctly identifies 226 of 338 targets (67\%) at 90\% precision (95\% CI: 86--93\%; 25 incorrect identifications, 86 abstentions).
\edit{}{We note that these edited profiles are much easier to identify than most pseudonymous accounts; we discuss this bias in \cref{ssec:method_eval}.}

We also evaluate the agent on two smaller Reddit datasets, academic users and career-discussion posters, where it achieves 25--52\% recall at 72--90\% precision; we report the full details in \cref{app:reddit_deanon}.

\paragraph{Anthropic Interviewer transcripts.}
Anthropic publicly released transcripts of 125 AI-conducted interviews with scientists~\citep{handa2025interviewer,anthropic2025interviewer}, in which participants described their use of AI in their research. The transcripts were partially redacted for participant privacy.
\citet{li2026agentic} showed that LLM agents with web search can match project descriptions in these transcripts to published papers;
specifically, they use a task decomposition approach~\citep{jones2024adversaries} to bypass safeguards by breaking the attack down into individual benign tasks.
They eventually ran their attack on 24 scientists conversations and recovered 6 identities. 
We attack the same dataset with our pipeline, \emph{without task decomposition}.
Concretely, we construct profiles from transcripts in a similar way as we do for Hacker News and Reddit users;
and then run the agentic search on 33 scientists that talked about their past research in the transcripts. 

The agent correctly identifies 9 out of 33 scientists at 82\% precision (2 wrong, 22 refusals or abstentions).
Here, we are uncertain about the true deanonymization rate, as we have to manually verify the matches. The agent claims to identify 11 scientists; but for 2 out of 11 we notice the transcript contradicts the identified academic profile in some ways.
We skip the anonymization step in this experiment: the transcripts by design do not contain platform handles or usernames, and we report deanonymization judged by a human, following the precedent set by \citet{li2026agentic}. Our judgment is that our agent finds 50\% more matches than \citet{li2026agentic}, but the exact accuracy in either of these experiments cannot be fully verified.

\paragraph{Limitations of the agentic approach.}
The agentic approach demonstrates that LLM-based deanonymization is a real threat, but it limits the science we can do.
First, ground-truth datasets are small because they require verified identity links.
Second, in privacy research the standard metric is a precision-recall curve; but our approach produces a single point estimate of precision and recall rather than a tunable tradeoff. Finally, the attack relies on opaque web search systems, making it difficult to isolate what the LLM agent contributes versus what the search engine embeddings contribute.

\section{A framework for scalable deanonymization with LLMs}
\label{sec:method}

\begin{figure*}[t]
    \centering
    \includegraphics[width=\textwidth, trim=0 50pt 0 0, clip]{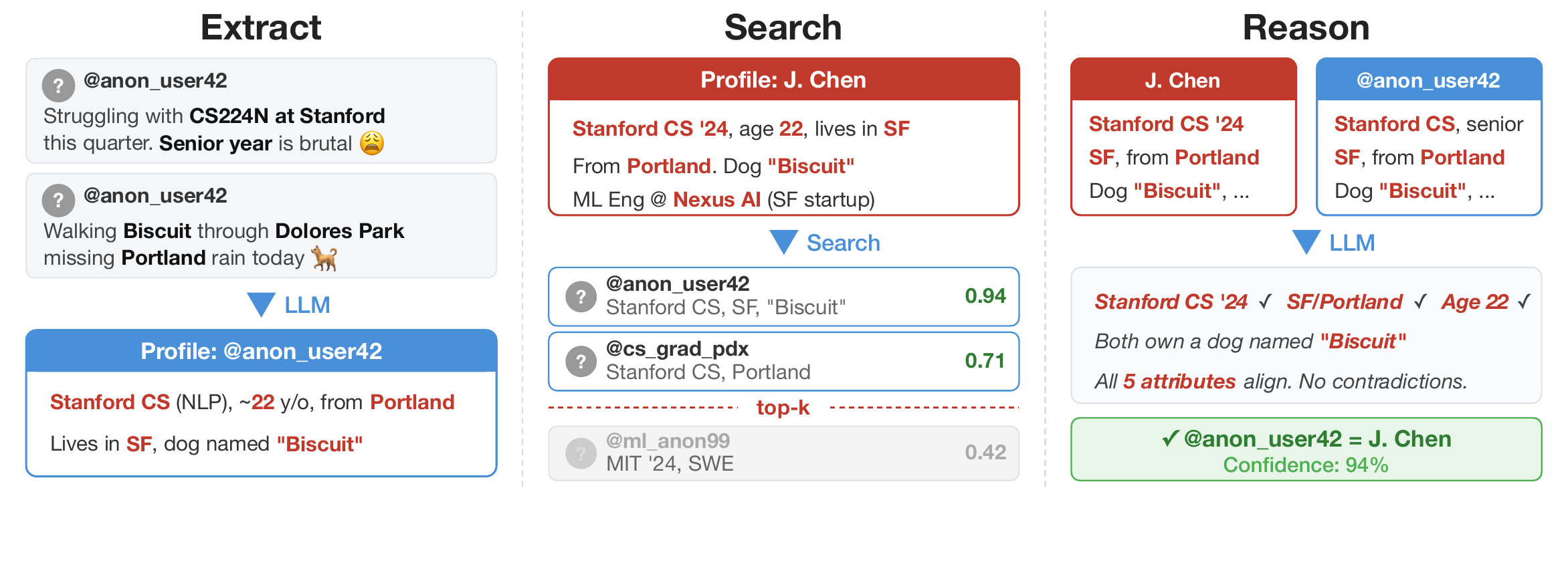}
    \caption{Overview of our framework for large-scale deanonymization. Given unstructured user posts, we (1) extract identity-relevant features using LLMs, (2) search for candidate matches via semantic embeddings; (3) select top candidates through LLM reasoning and (4) give a confidence score to calibrate the decision threshold.
    }
    \label{fig:esv_pipeline}
\end{figure*}

To better understand how LLMs deanonymize people, we need to explicitly state the different \edit{tasks that the LLM agent is solving}{steps of successful deanonymization}
For this reason, we introduce a modular framework that decomposes deanonymization into distinct stages, each of which can be run with or without an LLM.
\edit{This}{While non-agentic, this approach} 
lets us scale to larger datasets, trace precision-recall curves, and ablate the contribution of LLMs at each step.
Our framework is inspired by the seminal work of \citet{narayanan2008deanonymization}, whose attack serves as a ``classical'' baseline that we systematically augment with LLM components.

We first define our threat model and the ESRC deanonymization framework in \cref{ssec:method_threatmodel} and \cref{ssec:method_framework}, respectively.
In \cref{ssec:method_eval}, we then introduce an evaluation procedure
to quantify the effectiveness of LLM-based deanonymization for different settings
(identifying pseudonymous accounts from known identities in \cref{sec:study_hn}
and matching pseudonymous profiles in \crefrange{sec:study_movies}{sec:study_reddit}).

\subsection{Threat model}
\label{ssec:method_threatmodel}

We build on \cite{narayanan2008deanonymization}, who introduce large-scale deanonymization attacks by reconstructing \emph{micro-data}.
Micro-data is information at the level of an individual, such as ``gave Twilight a 5-star rating'', ``lives in Texas'', or ``never capitalizes sentences''.
This information alone may not be identifying, but it can identify a pseudonymous account by \emph{matching} their micro-data against a database of micro-data with known identities.
\citet{narayanan2008deanonymization} (hereafter termed the ``\nsattack'') demonstrate this by matching anonymized Netflix Prize ratings to public IMDb profiles based on movie rating micro-data.

\paragraph{Attacker goal.}
Our attacker's goal is to deanonymize pseudonymous online identities, which we model as the goal of matching profiles belonging to the same user across two sets.
Concretely, our attacker is given a \emph{query user} profile and a large set of \emph{candidate user} profiles.
Given those inputs, the attacker either returns a best-guess match of the query user in the candidate set or abstains.
The attacker's goal is to produce a correct guess if the query user has a corresponding candidate, and it should abstain if there is no matching candidate.

\paragraph{Controlling deanonymization difficulty.}
The difficulty of finding a match depends on two key factors: the number of candidates and the prior probability that the query user has a matching candidate.
Larger candidate pools obviously make the attack harder: given just two candidates (one of which is a correct match), even random guessing correctly matches about half of all queries. In contrast, identifying the correct user in 10k candidates requires a much stronger attacker.
The magnitude of the prior probability of a match influences the attacker's precision (i.e., if the prior is very low, then any non-abstaining guess is likely false \emph{a priori}).

For most of this paper, we will fix these parameters by using a fixed candidate pool for each setting and assuming a best-case scenario where every query user has a true match in the candidate set (i.e., the prior of a match existing is 100\%).
For full generality, we perform a detailed analysis in which we vary these two difficulty parameters in \Cref{ssec:case_reddit_ablation}.

\subsection{The ESRC deanonymization framework}
\label{ssec:method_framework}

Our deanonymization framework follows the structure of the seminal attack by \citet{narayanan2008deanonymization}, augmented with LLM components. \Cref{fig:esv_pipeline} illustrates how LLMs make each part of the pipeline easier.
Concretely, all our experiments involve the following steps: 
\begin{enumerate}
    \item \textbf{Extract}: The attack first needs to extract micro-data for query and candidate users.
    The original \nsattack directly receives structured micro-data in the form of Netflix and IMDb rating vectors as input. For arbitrary online profiles, we use LLMs to extract semi-structured summaries from unstructured posts and comments. As a result, the LLM-augmented attack may be able to rely on much more information.
    
    \item \textbf{Search}: Given micro-data for the query and candidate users, the second stage of the attack searches for the most likely matches.
    In the \nsattack, this is done with a simple (weighted) vector similarity search over feature vectors (i.e., the candidate with the highest vector similarity is returned).
    For our method, we do this by performing a nearest-neighbor search over \emph{LLM embeddings} of the extracted summaries.

    \item \textbf{Reason:}
    LLMs allow us to select from a shortlist of multiple top matches returned by the Search step, whereas the \nsattack could only return the single best match by vector similarity.
    Additionally, we can first use a cheaper LLM to select the most likely candidate, then use a more expensive model with advanced reasoning capabilities to verify the selection.

    \item \textbf{Calibrate:}
    \edit{Finally, we need to calibrate the tradeoff between precision---how many of our predictions are correct---and recall---how many true matches we identify.}{Finally, we need to calibrate the tradeoff between precision (how many of our predictions are correct) and recall (how many true matches we identify).}
    The \nsattack does this by thresholding the similarity score of the top match to control the attack's confidence (i.e., if the top match's similarity is above a given threshold, the attacker guesses this match, and otherwise abstains).
    Similarly, we can use the embedding similarity score from the Search step, or an LLM-produced confidence score from the Reason step, or ratings computed by sorting matches via pairwise LLM comparisons.
    We use these confidence scores to trace precision-recall curves.
\end{enumerate}

\subsection{Experimental Setup}
\label{ssec:method_eval}

To enable quantitative evaluations of the benefits of LLMs at scale, we instantiate our deanonymization framework in settings with verifiable ground truth.
To obtain this ground truth, we consider two approaches:

\begin{enumerate}
    \item \emph{Synthetically anonymized matches.}
We search for user profiles that explicitly link themselves to other platforms (e.g., a Hacker News user that posts their LinkedIn profile). Since these profiles trivially expose links, we carefully sanitize them by removing all identifying information.
\edit{}{Text sanitization is a well-studied problem~\citep{morris2022unsupervised, Staab2025LanguageMA}, though recent work shows that even sophisticated methods leave semantic signals that enable re-identification~\citep{xin2025false}; the point of our attacks is to exploit precisely the signals that mimic the traces that a genuinely anonymous user would leave in their profile.}
\item \emph{Semantic splits for arbitrary profiles.}
In practice, we found only few users who explicitly link their pseudonymous account to other platforms.
Hence, as a second option, we split a single user's profile into two semantically distinct parts (e.g., temporally). Those splits provide the strongest source of ground truth information at the cost of being less realistic.
\end{enumerate}

Any deanonymization setup with ground truth introduces distributional biases.
In our cross-platform datasets, the profiles are likely easier to deanonymize than an average profile: the very fact that ground truth exists implies that the user may not have cared about anonymity in the first place.
Similarly, two split-profiles of a single user are inherently alike, whereas two pseudonymous accounts of the same person (e.g., an official and a pseudonymous alt account) might expose more heterogeneous micro-data.

We thus focus on methods with very few false positives.
Due to the aforementioned biases, our evaluation on ground truth settings might overestimate an attack's recall (how many users are successfully deanonymized).
But we argue that the false positive rate (i.e., the likelihood of a wrong guess instead of abstaining) does transfer to real-world attacks; there is no reason why a well-calibrated attacker should make more wrong guesses on the average internet profile than on our ground truth profiles.
Therefore, similarly to \cite{narayanan2008deanonymization}, we focus on attacks with a very low false positive rate.
This ensures that whenever our attacks do return a guess, the guess is likely to be correct---independent of the setting.

Concretely, we report \emph{\metric} for high precision values.
We hence define \emph{recall} (or \emph{True Positive Rate (TPR)}) as the fraction of matchable users (i.e., users with a corresponding match in the candidate pool) that the attacker guesses correctly,
and \emph{precision} as the fraction of all non-abstentions that are correct.
We vary the confidence threshold at which we output a match
(derived from LLM embedding similarity or the Calibrate stage of the pipeline)
to get a precision-recall curve.

\section{Linking profiles across platforms: Hacker News and LinkedIn}
\label{sec:study_hn}

We begin our evaluation with cross-platform matching between LinkedIn and Hacker News. This setting reflects a plausible real-world attack where an adversary knows a person's real identity (e.g., from LinkedIn) and seeks their pseudonymous account on a different platform.

\subsection{Dataset}
We collect 987 LinkedIn profiles linked to 995 Hacker News (HN) accounts (ground truth is established by users who posted their LinkedIn URL in their HN bio), drawn from a candidate pool of approximately 89,000 active HN users. Eight LinkedIn profiles are linked to multiple HN accounts that shared the same LinkedIn URL. We identified four additional HN alt accounts using strong evidence such as matching names and companies. We count a match as correct if any of the linked HN accounts is returned. Every query has a true match in the candidate set. The LinkedIn side represents the known identity with real professional profiles. The HN side serves as the anonymized target: as in \Cref{sec:study_agents}, we remove names, URLs, and other direct identifiers from bios using an LLM to prevent trivial matching (see \cref{app:anonymization} for our complete anonymization procedure). The task is to match a LinkedIn profile with the corresponding LLM-anonymized HN account.

\subsection{Attack instantiation}

\paragraph{Extract: biographical profiles.}
We use an LLM to summarize each user's HN activity (comments, stories, and bio) into a biographical profile. On the LinkedIn side, we use an LLM to summarize the data that users reveal about themselves, such as job positions and school graduations.

\paragraph{Search: embedding similarity.}
We embed all 89{,}000 candidate profiles into a nearest-neighbor search index using Gemini embeddings~\citep{lee2025gemini} and FAISS~\citep{douze2025faiss} with cosine similarity. For each query, we embed their biographical profile and retrieve the top candidates.

\paragraph{Reason: two-stage selection and verification.}
Embedding similarity is effective at narrowing down the candidate pool to the most similar candidates, but the correct match will often not be ranked first. We therefore select the top-100 candidates by embedding similarity, then apply a two-stage reasoning process (illustrated in \cref{fig:matching_illustration} for the movie setting of \cref{sec:study_movies}). First, Grok 4.1 Fast~\citep{xai2025grok41fast} selects the most likely match from the shortlist or abstains. Then, if a candidate was selected, GPT-5.2~\citep{openai2025gpt52_system_card} evaluates the selected pair with either low or high reasoning effort. By only using a more capable model with advanced reasoning on a single promising match per query, we substantially reduce cost.

\paragraph{Calibrate: LLM confidence scores.}
We use the verification-stage confidence as the calibration score for precision-recall curves. For embedding-only and Narayanan baseline results, we use the gap between the top-2 candidates' similarities---a large gap indicates the top candidate stands out clearly, making it more likely correct. \edit{The model outputs confidence via function calling}{The model outputs a numerical confidence score as a structured field in its function-call response},
and we threshold it to trade off precision and recall, as shown in \cref{fig:linkedin_hn_pr}.

\paragraph{Baseline.}
We adapt the \nsattack to professional attributes: programming languages, cities, companies, schools, job roles, and seniority levels. Each user is represented as a binary vector over 332 attributes, and we rank candidates by rarity-weighted Jaccard similarity (see \cref{app:narayanan_baseline} for details).

\subsection{Results}

\begin{figure*}[t]
\centering
\begin{subfigure}[t]{\fighalf}
\centering
\includegraphics[width=\textwidth]{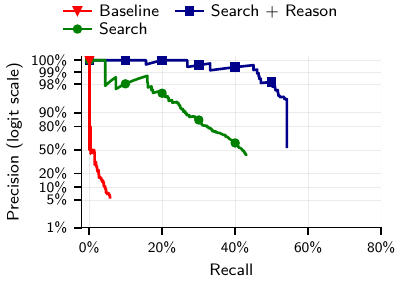}
\caption{Precision-recall curves.}
\label{fig:linkedin_hn_pr}
\end{subfigure}%
\hfill\begin{subfigure}[t]{\fighalf}
\centering
\includegraphics[width=\textwidth]{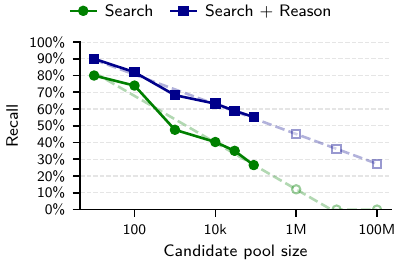}
\caption{Scaling with candidate pool size.}
\label{fig:scaling_analysis}
\end{subfigure}
\caption{LinkedIn to Hacker News matching. (a) Precision-recall curves comparing methods: LLM-based embeddings outperform the \nsattack baseline, and LLM selection from the top-100 candidates further improves performance. (b) Matching recall @90\%precision by candidate pool size. Dashed lines show log-linear extrapolation to larger pools.}
\label{fig:linkedin_hn_results}
\end{figure*}

\paragraph{LLM-based embeddings outperform handcrafted features.}
\Cref{fig:linkedin_hn_pr} shows the precision-recall tradeoff. The embedding-based approach substantially outperforms the \nsattack on all metrics; the \nsattack correctly ranks only 5.1\% of candidates as the top-1 match, and thresholding the top-2 similarity gap to improve precision quickly drives recall to near zero.

\paragraph{LLM reasoning further improves matching.}
As shown in \cref{fig:linkedin_hn_pr}, adding the Reason step substantially outperforms pure embedding search in both precision and recall. The ability of LLMs to select from a number of candidates and use reasoning complements embedding-based retrieval.

\paragraph{High reasoning effort helps at high precision.}
As shown in \cref{tab:hn_linkedin_recall_at_precision}, high reasoning effort significantly outperforms low reasoning only at 99\% precision (45.1\% vs 36.0\% recall). At lower precision thresholds, the two reasoning modes perform comparably.

\begin{table}[htbp]
\centering
\small
\caption{Recall at different precision thresholds for HN-LinkedIn cross-dataset user matching (987 queries). Reason uses GPT-5.2 with low or high reasoning effort. 95\% Wilson confidence intervals shown.}
\label{tab:hn_linkedin_recall_at_precision}
\begin{tabular}{@{} lccc @{}}
\toprule
Method & 90\% Prec & 98\% Prec & 99\% Prec \\
\midrule
Narayanan baseline & 0.1\% & 0.1\% & 0.1\% \\
 & \footnotesize (0.0--0.6) & \footnotesize (0.0--0.6) & \footnotesize (0.0--0.6) \\[0.3em]
Search (embedding) & 26.3\% & 16.1\% & 4.4\% \\
 & \footnotesize (23.7--29.2) & \footnotesize (13.9--18.5) & \footnotesize (3.3--5.9) \\[0.3em]
Reason (low) & \textbf{55.0\%} & \textbf{44.8\%} & 36.0\% \\
 & \footnotesize (51.9--58.0) & \footnotesize (41.8--47.9) & \footnotesize (33.1--39.0) \\[0.3em]
Reason (high) & \textbf{54.2\%} & \textbf{50.0\%} & \textbf{45.1\%} \\
 & \footnotesize (51.1--57.2) & \footnotesize (46.9--53.0) & \footnotesize (42.1--48.2) \\
\bottomrule
\end{tabular}
\end{table}

\paragraph{Reasoning scales to larger candidate pools.}
We evaluate Search and Reason as the pool grows from 1k to 89k users. As shown in \cref{fig:scaling_analysis}, both methods degrade log-linearly with pool size, but Reason is substantially more robust. At 90\% precision, Reason achieves 68.3\% recall at 1k candidates and retains 55.2\% at 89k---a loss of only 13 percentage points across nearly two orders of magnitude. Search drops more steeply, from 47.6\% to 26.6\% over the same range. The gap widens at high precision: at 89k candidates and 99\% precision, Reason still achieves 33.0\% recall while Search collapses to 4.2\%. Log-linear extrapolation to 1M candidates projects Reason at approximately 45\% recall (90\% precision), compared to roughly 12\% for Search; at 100M candidates, Reason would still retain an estimated 27\% recall while Search falls to zero (see \cref{app:scaling} for details).

\section{Linking users across Reddit communities}
\label{sec:study_movies}

The original Netflix Prize attack was designed for connecting cross-platform movie ratings, not the cross-platform professional setting of \cref{sec:study_hn}. We thus construct a more direct comparison using movie-related Reddit activity. This provides a clean comparison: both methods operate on the same micro-data features (subjective movie reviews) but differ in how they represent users---we use transformer-derived text embeddings, while \citet{narayanan2008deanonymization} relied on hand-crafted numerical features.

\subsection{Constructing the community-split dataset}

Although the Hacker News--LinkedIn setting from \cref{sec:study_hn} demonstrates cross-platform deanonymization, its scale is limited by verifiable ground truth.
Profiles for which we can confidently infer ground truth matches are rare---they require users to explicitly link accounts or directly disclose their identity on both profiles. This is a general problem when evaluating deanonymization, since the anonymity that motivates the problem also obstructs collecting ground truth data.

We solve this problem by splitting Reddit profiles: partitioning each user's activity into separate query and candidate profiles. This yields thousands of profile pairs with exact ground truth. In this section, we split by community; in \cref{sec:study_reddit}, we split temporally.

This setup mimics a situation in which a user maintains pseudonymous accounts on different platforms discussing similar topics. In our case, the communities discuss either mainstream movies (r/movies) or more niche movie categories. This also helps understand the relationship between how much micro-data someone shares and how identifiable they become. Unlike most online activity, movie discussions offer an intuitive discrete metric: the number of movies discussed and the overlap of shared movies across datasets.

Our dataset consists of movie discussion communities (``subreddits''), restricted to comments and submissions made in 2024. We collect all users who post in both r/movies (the largest general movie discussion community) and at least one of five smaller specialized communities (r/horror, r/MovieSuggestions, r/Letterboxd, r/TrueFilm, and r/MovieDetails); we refer to the union of these five as the \emph{alternative movie communities}. We treat each user's r/movies activity as the query profile and their activity in the alternative communities as the candidate profile; the candidate set contains all 9{,}781 users, so every query has a true match in the candidate set. Users share an average of 2.5 movie titles between the two splits (determined via fuzzy string matching of titles mentioned in posts), with 48\% sharing only a single movie. We retain only users with at least one overlapping title, since no method restricted to movie-preference micro-data can match users with zero overlap.

\subsection{Attack instantiation}

\paragraph{Extract: structured movie reviews.}
We use LLMs to convert unstructured Reddit submissions---which may contain discussions, recommendations, or casual mentions---into structured movie reviews with movie titles, ratings, and dates. Ratings are typically estimated by the LLM based on sentiment. The resulting user profiles consist solely of movie reviews and contain no other micro-data.

\paragraph{Search: embedding similarity.}
As in \cref{sec:study_hn}, we encode the extracted review profiles into dense embeddings using Gemini embeddings and FAISS, and retrieve the top-100 nearest neighbors by cosine similarity on normalized vectors.

\begin{figure}[t]
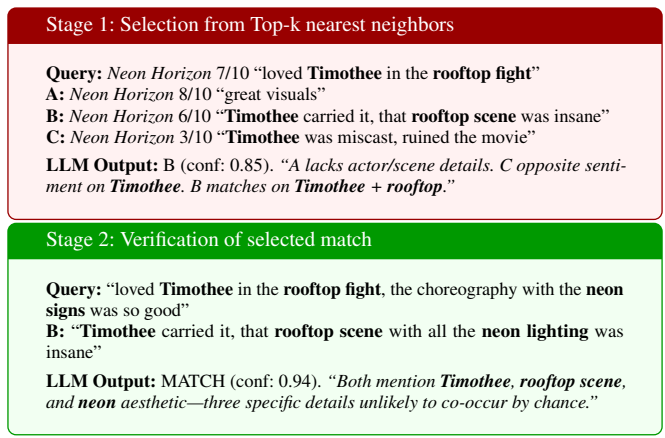

\centering
\begin{minipage}[t]{\fighalf}
\begin{tcolorbox}[colback=red!5, colframe=red!60!black, boxrule=0.5pt, title={\footnotesize Stage 1: Selection from Top-k nearest neighbors}]
\scriptsize
\textbf{Query:} \textit{Neon Horizon} 7/10 ``loved \textbf{Timothee} in the \textbf{rooftop fight}''

\textbf{A:} \textit{Neon Horizon} 8/10 ``great visuals''

\textbf{B:} \textit{Neon Horizon} 6/10 ``\textbf{Timothee} carried it, that \textbf{rooftop scene} was insane''

\textbf{C:} \textit{Neon Horizon} 3/10 ``\textbf{Timothee} was miscast, ruined the movie''

\vspace{0.1cm}
\textbf{LLM Output:} B (conf: 0.85). \textit{``A lacks actor/scene details. C opposite sentiment on \textbf{Timothee}. B matches on \textbf{Timothee} + \textbf{rooftop}.''}
\end{tcolorbox}
\end{minipage}%
\hfill
\begin{minipage}[t]{\fighalf}
\begin{tcolorbox}[colback=green!5, colframe=green!60!black, boxrule=0.5pt, title={\footnotesize Stage 2: Verification of selected match}]
\scriptsize
\textbf{Query:} ``loved \textbf{Timothee} in the \textbf{rooftop fight}, the choreography with the \textbf{neon signs} was so good''

\textbf{B:} ``\textbf{Timothee} carried it, that \textbf{rooftop scene} with all the \textbf{neon lighting} was insane''

\vspace{0.1cm}
\textbf{LLM Output:} MATCH (conf: 0.94). \textit{``Both mention \textbf{Timothee}, \textbf{rooftop scene}, and \textbf{neon} aesthetic---three specific details unlikely to co-occur by chance.''}
\end{tcolorbox}
\end{minipage}
\caption{Two-stage matching of movie reviews with simplified synthetic user profiles. The first LLM selects from top-K embedding similarity candidates; the second LLM verifies the selected match using full profile text.
Bold text highlights profile details used by the LLM.}
\label{fig:matching_illustration}
\end{figure}

\paragraph{Reason: two-stage selection and verification.}
In contrast to \cref{sec:study_hn}, where we directly select from embedding neighbors, here we use a two-stage approach (\cref{fig:matching_illustration}). First, we select from the top-K candidates using only reviews with matching movie titles between query and candidates. Then, we verify the top match using full profile text. This reduces cost---selection sees shorter inputs while verification runs only on the top candidate---and allows using a more capable model for verification. We use GPT-5.2~\citep{openai2025gpt52_system_card} with high reasoning effort.

\begin{figure*}[t]
\centering
\begin{subfigure}[t]{\fighalf}
\centering
\includegraphics[width=\textwidth]{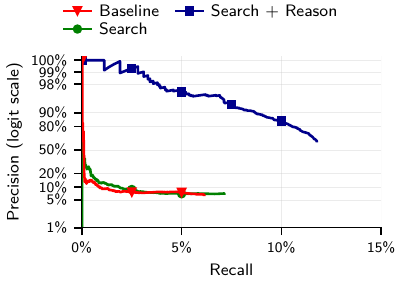}
\caption{Precision-recall curves for Reddit movies matching.}
\label{fig:pr_curve}
\end{subfigure}%
\hfill
\begin{subfigure}[t]{\fighalf}
\centering
\includegraphics[width=\textwidth]{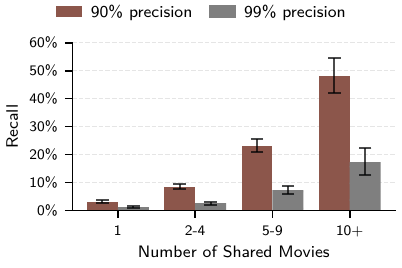}
\caption{Recall by number of shared movies at different precision levels.}
\label{fig:recall_by_shared_movies}
\end{subfigure}
\caption{(a) Precision-recall curves comparing methods for Reddit movie matching. LLM reasoning outperforms both the embedding gap baseline and the \nsattack.
(b) Recall at precision thresholds by number of shared movies, using GPT-5.2 high reasoning. Users who share more movies are substantially easier to identify. See \cref{tab:recall_by_shared_movies} for detailed breakdown.}
\label{fig:reddit_matching_results}
\end{figure*}

\paragraph{Calibrate: LLM confidence scores.}
We use the verification-stage confidence as the calibration score for precision-recall curves; the selection-stage confidence is used only internally to pick the candidate passed to verification. \edit{The model outputs confidence via function calling}{The model outputs a numerical confidence score as a structured field in its function-call response}, and we threshold it to trade off precision and recall. As a simpler alternative, we also evaluate using the gap between the top-2 embedding similarities as the confidence score---a large gap indicates the top candidate stands out clearly. As shown in \cref{fig:pr_curve}, LLM confidence substantially outperforms this gap-based baseline.

\paragraph{Baseline.}
Both the LLM-based method and the classical baseline consume the same LLM-extracted (title, inferred rating, timestamp) tuples; they differ only in the matching procedure. As a classical baseline, we adapt the \nsattack: each user is represented as a vector of movie ratings (analogous to the Netflix Prize setting), and candidates are ranked by rarity-weighted similarity over shared movies, where popular movies are downweighted to emphasize distinctive preferences. See \cref{app:narayanan_baseline} for details.

\subsection{Results}
\paragraph{Higher reasoning effort substantially improves matching.}
We evaluate whether setting the reasoning effort to low or high affects the matching performance. High reasoning significantly outperforms low reasoning across all thresholds: at 90\% precision, high reasoning achieves 8.5\% recall compared to 5.2\% for low reasoning (a relative improvement of 63\%). At 99\% precision, high reasoning achieves 2.8\% recall versus 1.4\% for low reasoning (a relative improvement of 100\%). These differences are statistically significant ($p < 0.05$; see \cref{tab:recall_at_precision}). Increased reasoning effort (test-time compute) substantially improves deanonymization success.

\paragraph{Users who share more content are substantially easier to identify.}
As shown in \cref{fig:recall_by_shared_movies}, recall increases sharply with the number of shared movies: at 90\% precision, recall rises from 3.1\% for users with one shared movie to 48.1\% for users with ten or more.

\paragraph{Sorting matches does not improve over LLM confidence scores.}
In \cref{sec:study_reddit}, we introduce sorting matches with an LLM by perceived match likelihood as an alternative calibration method; we compare both approaches in \cref{app:tournament_movies}. We attempted to apply it to this setting on top of our two-stage selection and verification pipeline. However, sorting was unable to improve over the LLM confidence scores from the verification stage.

\section{Scaling LLM deanonymization on temporally split profiles
}
\label{sec:study_reddit}

As in the previous section, we split Reddit profiles to obtain large-scale ground truth. Here we split temporally rather than by community: we partition each user's comments into a before (query) and after (candidate) set, separated by a one-year gap. The one-year gap ensures that the matching task is non-trivial, as users' discussions of current events and short-term interests differ across splits.
The resulting dataset allows us to study LLM-based deanonymization in depth and at scale. We also explore a different calibration approach based on pairwise comparisons rather than confidence scores,
and we evaluate the attacks' robustness to difficulty parameters of our attack model.

\subsection{Constructing the temporal-split dataset}

We construct 10{,}000 query and candidate profiles as follows.
We begin with 5{,}000 users, splitting each's comments into a query profile and a candidate profile.
To increase difficulty, we add 5{,}000 candidate distractors: candidate profiles of users who appear only in the candidate pool, with no corresponding query.
This collection of 5{,}000 query and 10{,}000 candidate profiles form the core matching task in \cref{ssec:case_reddit_main},
where every query has a match in the candidate set.
To evaluate attack models where queries can be non-matchable (\cref{ssec:case_reddit_ablation}),
we further add 5{,}000 query distractors: additional users who appear only in the query set, with no true match in the candidate pool.
We construct two such datasets independently: a development set for tuning our attacks and a held-out test set for our final evaluation.
All results presented in this section use the test set.

We apply several filters to ensure the matching task is meaningful and non-trivial.
Our filters ensure that all 15{,}000 users are sufficiently active, but not too active.
This yields split-profiles with sufficient micro-data while discarding bots.
We also discard all comments within a one-year window around the split time of each user, so that the two split profiles do not share contemporaneous discussions of current events or short-term life circumstances.
See \cref{app:reddit_pipeline} for full details.

The temporal gap and difference in content between the split-profiles resembles what an attacker might face when linking an abandoned account to a newly created one or matching a user's main account to their alt-account.
This requires identifying stable micro-data (e.g., user characteristics, interests, writing style, demographics) from hundreds of comments.

\subsection{Attack instantiation}

\paragraph{Extract: comment summaries.}
To extract micro-data features, we use LLMs to filter and summarize the comments of each split-profile.
We first apply a two-stage relevance filter: a heuristic pre-filter removes empty and deleted comments, very short responses, and pure URLs. Then, we prompt Gemini 3 Flash to label each of the remaining comments as relevant or generic. We discard generic comments and feed the remaining relevant ones to Gemini 3 Pro~\cite{google2025gemini3pro}.
The model generates a comma-separated list of the most important details,
resulting in semi-structured micro-data.
We discard both the query and candidate profile of users with zero comments after filtering (2 users)
or if the LLM refuses to generate summaries due to inappropriate content (83 users).
See \cref{app:prompts} for prompts and examples.

\paragraph{Search: cosine similarity over embedding.}
Similar to the previous two sections,
we perform a nearest neighbor search (in terms of cosine similarity)
between LLM embeddings of the extracted summaries.
We generate the embeddings using OpenAI's \texttt{text-embedding-3-large} model.

The Extract and Search steps above yield a base attack:
for each query, we return the candidate with the highest cosine similarity, using the similarity itself as the confidence for calibration.
To justify the Reason and Calibrate steps, we investigate this base attack in more detail for the current setting.
As detailed in \cref{app:method_justification}, the attack has two key issues.
First, the embedding similarity successfully narrows the candidate pool but fails to reliably identify a single best guess.
Second, embedding similarity is a poor confidence measure at high precision,
such that high-similarity query-candidate pairs are not reliably correct matches.
This motivates the design of the Reason and Calibrate steps.

\paragraph{Reason: LLM selection.}
As in the previous sections, we use LLMs to select the best match from the top-15 candidates (in terms of embedding similarity).
We set $k=15$ since we found that, on the training set, 80\% of true matches fall within this range.\footnote{
The numbers are similar for the held-out test set; see \cref{fig:method_justification}.
}
For each query, we give the 15 highest-scoring candidate summaries to Gemini~3~Pro, and ask it to select which best matches the query user's summary.

\paragraph{Calibrate: sorting matches via tournament.}
Since similarity scores are poorly calibrated confidence measures, we sort all proposed query-candidate matches from most to least plausible using pairwise LLM comparisons.
Concretely, we implement a Swiss-system tournament over all selected query-candidate pairs (\cref{alg:sorting}), using the smaller GPT-5-mini \cite{singh2025openaigpt5card} model for efficiency.
In each round, pairs of matches are compared head-to-head: the LLM sees two query-candidate pairs and judges which is a more plausible match.
After each round, we update Bradley-Terry ratings~\citep{bradley1952rank} based on the comparison outcomes.
We run 15 rounds with approximately 2{,}500 comparisons per round, and output matches sorted by their final rating.

\begin{algorithm}[t]
\caption{LLM-based confidence sorting 
}
\label{alg:sorting}
\begin{algorithmic}[1]
\Require Set of query-candidate pairs $\{(q_i, c_i)\}$ from the Reason step
\State Initialize Bradley-Terry~\citep{bradley1952rank} ratings $r_i$ for all pairs
\For{round $= 1$ to $N$}
    \State Pair up matches by similar rating (Swiss-system matching)
    \State For each pair $(q_i, c_i)$ vs.\ $(q_j, c_j)$: LLM judges which is the more plausible match
    \State Update ratings $r_i, r_j$ based on comparison outcome
\EndFor
\State \Return pairs sorted by final rating $r_i$ (descending)
\end{algorithmic}
\end{algorithm}

This sorting procedure avoids the quadratic cost of comparing all proposed query-candidate pairs, and we find it to be an effective confidence measure.
However, sorting depends on interactions between query users; as such, it explicitly requires a large set of queries to be effective.
This requirement makes sorting infeasible for attackers who aim to deanonymize a single user, but it is well-suited for large-scale attacks.

\paragraph{Baseline: \nsattack on subreddit participation.}
Each user is represented as a binary vector indicating which subreddits the user posted a comment in.
We then directly instantiate the \nsattack using this structured micro-data; see \cref{app:narayanan_baseline} for details.

\subsection{Results}
\label{ssec:case_reddit_main}
We evaluate our LLM augmentations in the same attacker model as before, that is, with a fixed candidate pool that contains a ground truth match for every query user.
Concretely, we run the attack for the 5{,}000 query profiles with a true match in the candidate set, and we fix the candidate pool to all 10{,}000 candidate profiles.

\begin{figure*}[t]
    \centering
    \begin{subfigure}[t]{\figsixcol}
        \centering
        \includegraphics[width=\textwidth]{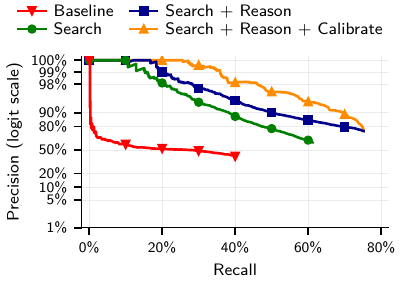}
        \caption{Full Precision-Recall curves.}
        \label{fig:reddit_main_pr}
    \end{subfigure}
    \hfill
    \begin{subfigure}[t]{\figsixcol}
        \centering
        \includegraphics[width=\textwidth]{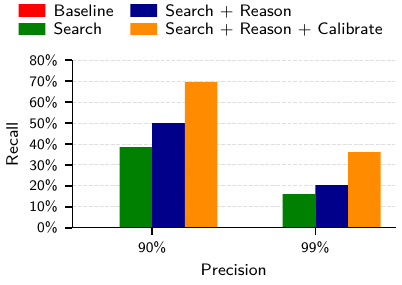}
        \caption{\metric of the classical baseline (close to 0\%) compared to LLM-based attacks.}
        \label{fig:reddit_main_bars}
    \end{subfigure}
    \caption{\textbf{Classical attacks fail to deanonymize split Reddit profiles, while LLM-based attacks are highly effective.}
    We compare a classical baseline that mimics the \nsattack to LLM deanonymization.
    (a) The precision of classical attacks drops very fast, explaining its low recall. In contrast, the precision of LLM-based attacks decays more gracefully as the attacker makes more guesses.
    (b) The classical attack almost fails completely even at moderately low precision. In contrast, even the simplest LLM attack (Search) achieves non-trivial recall at low precision, and extending it with Reason and Calibrate steps doubles \metric[0.99].
    }
    \label{fig:reddit_main}
\end{figure*}

\paragraph{LLM-based extraction and search outperforms the classical baseline.}
As shown in \cref{fig:reddit_main_bars}, the classical attack, similar to the \nsattack, fails to achieve non-trivial recall at only 90\% precision.
In contrast, even our simplest attack achieves non-trivial recall at all precision levels.%

\paragraph{LLMs are good at picking the correct match from a small set of candidates.}
Embedding similarity effectively narrows the candidate set: for about 80\% of queries, the true match ranks among the top 15 candidates (see \cref{fig:topk_recall}). 
Using an LLM to select the best candidate from the top 15, we recover many of these missed matches, increasing recall at high precision (``Search + Reason'' in \cref{fig:reddit_main_pr}).

\paragraph{LLMs can prioritize more likely matches.}
The results in \cref{fig:reddit_main} confirm our hypothesis that embedding similarity is a subpar confidence measure:
Adding tournament-sorting significantly boosts recall across all precision values.
In particular, our full  attack (``Search + Reason + Confidence'')
reaches a recall closer to the best-possible value (80\% imposed through the Reason step) at 90\% precision,
and it still deanonymizes a third of all users at 99\% precision.

\subsection{Comparing difficulty parameters of our attack model}
\label{ssec:case_reddit_ablation}

We now study the two key factors that determine the difficulty of our attack model (\cref{ssec:method_threatmodel}).
As a reminder, a larger candidate pool makes it more challenging to find a correct match, and a lower a priori likelihood of there being a matching candidate requires more abstentions.
Since these factors are typically unknown in practice, we investigate their full range in the following.
This allows us to extrapolate how well our attacks can work in various real-world settings.

\paragraph{Setup.}
We rerun the baseline and our strongest LLM-based attack (Search + Reason + Calibrate) on candidate pools of various sizes.
For each size, we sample subsets of the full 10k candidate profiles while ensuring that smaller pools are included in the larger ones (and the true match is always present).
Since this procedure is random, we repeat it over five candidate set draws for the baseline and three draws for the (more expensive) LLM-based attack.

We further linearly extrapolate attack success to much larger candidate pools.
Although our candidate pool of 10k users is reasonably large, platforms such as Reddit likely yield in the order of a million candidates, even after pre-filtering.
Concretely, we fit a linear model to \metric as a function of $\log_{10}(K)$,
where $K$ is the size of the candidate set.
To avoid overestimating attack success, we omit values for $K=10$.
This extrapolation paints a coarse picture of attack success in internet-scale settings.

For the second difficulty parameter, let $\matchfrac \in [0, 1]$ be the a priori probability that a query user has a matching candidate.
Empirically, if $M$ query-profiles are matchable and $N$ profiles are not, $\matchfrac = M / (M + N)$.

We can calculate \metric for all values of $\matchfrac$ post-hoc.
First, notice that $\matchfrac$ only affects precision.
Precision decreases through two types of error rates, depending on whether a query user has a true match in the candidate set:
\begin{itemize}
    \item \textbf{False Match Rate (\fmr)}: the fraction of \emph{matchable} queries for which the attacker returns a wrong guess.
    \item \textbf{False Positive Identification Rate (\fpir)}: the fraction of \emph{non-matchable} queries for which the attacker returns any guess (i.e., does not abstain).
\end{itemize}
As before, \tpr{} (or recall) is the fraction of matchable queries for which the attacker returns the correct guess.

Precision is a function of those three rates and the fraction of matchable users $\matchfrac$ (see \cref{app:precision_pi} for the derivation):
\begin{equation*}
    \precision(\pi) = \frac{
    \matchfrac \cdot \tpr
}{
    \matchfrac \cdot \tpr
    + \matchfrac \cdot \fmr
    + (1 - \matchfrac) \cdot \fpir
} .
\end{equation*}

Crucially, the \tpr{} and \fmr{} only depend on the matchable queries, while the \fpir{} only depends on the non-matchable queries.
This allows us to simulate different attack models:
we first estimate the three rates and then use the reformulated precision as a plug-in estimator to calculate \metric for multiple values of $\matchfrac$ post-hoc.
Concretely, we use the full set of all 10k candidate profiles,
but we run our attack on both the 5k query profiles with a true match in the candidate set and the additional 5k non-matchable query profiles.

We only evaluate the effects of the fraction of matchable users $\matchfrac$ for the Search and Search + Reason attacks.
Since the tournament-based Calibration step correlates queries,
we cannot calculate \metric post-hoc for our strongest attack.
Moreover, the resulting metric has high variance for small $\matchfrac$ values.
Thus, we would need to rerun the attack many times for accurate estimates,
which is prohibitively expensive.

\begin{figure*}[t]
    \centering
    \begin{subfigure}[t]{\figsixcol}
        \centering
        \includegraphics[width=\textwidth]{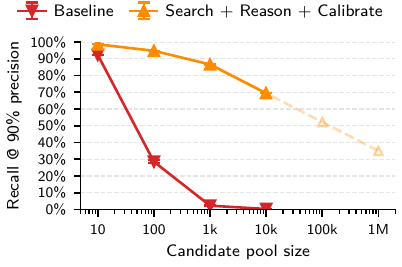}
        \caption{Effects of candidate set size}
        \label{fig:reddit_ablations_candidates}
    \end{subfigure}
    \hfill
    \begin{subfigure}[t]{\figsixcol}
        \centering
        \includegraphics[width=\textwidth]{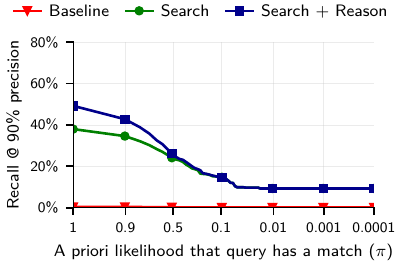}
        \caption{Effects of the a priori likelihood that a query has a match ($\matchfrac$)}
        \label{fig:reddit_ablations_matchfrac}
    \end{subfigure}
    \caption{\textbf{LLM deanonymization scales to more challenging attack models, while classical methods fail.}
    (a) The strongest LLM-based attack (Search + Reason + Calibrate) might remain effective for very large candidate pools, while the classical baseline fails to deanonymize any users given much smaller candidate pools. We log-linearly extrapolate the recall from sizes 100, 1k, and 10k.
    (b) LLM-based attacks gracefully degrade as matchable users become more rare, converging to around 9\% recall even if only one in 10k queries has a possible match. In contrast, the classical baseline fails to deanonymize users even if all of them are matchable ($\matchfrac=1$).
    See \cref{app:results} for more precision levels.
    }
    \label{fig:reddit_ablations}
\end{figure*}

\paragraph{LLM-based attacks extrapolate to internet-scale datasets.}
As seen in \cref{fig:reddit_ablations_candidates}, the recall of our LLM-base attack decreases roughly linearly in the scale of the dataset.
Extrapolating to one million candidates, the LLM-based attack still yields about 35\% recall at 90\% precision.
In contrast, the classical attack achieves a lower recall even for just a hundred candidates.
We note that this is a coarse extrapolation, which should be read with large margins of error.
Nevertheless, we conjecture that LLM deanonymization scales to internet-scale candidate pools with non-trivial success.

\paragraph{LLMs confidently deanonymize many users even if true matches are extremely rare.}
\Cref{fig:reddit_ablations_matchfrac} shows a surprising trend:
LLM-based attacks consistently achieve a recall of at least 9\% at 90\% precision---even if the probability of a query having any match is only one in 10{,}000 ($\matchfrac = 0.0001$).
We hence conjecture that, even in settings where almost no users can be deanonymized, LLM-based attacks are reasonably likely to find a correct match for the few users that are identifiable.
This demonstrates how LLM-based deanonymization still works in hard real-world settings.

\section{Related work}
\label{sec:related_work}

\paragraph{Deanonymization attacks predate LLMs. \edit{but were limited to structured data or high manual effort}{}}
\citet{narayanan2008deanonymization} demonstrated that movie ratings can uniquely identify individuals across platforms: by matching Netflix Prize data against public IMDb profiles, they deanonymized users at scale using statistical techniques on structured micro-data.
Their work established the threat model we build upon, showing that even sparse, seemingly innocuous data can be identifying when matched against auxiliary datasets.
The same authors showed that social graph structure alone can deanonymize users by matching connection patterns across networks~\citep{narayanan2009deanonymizing}.
\citet{wondracek2010practical} exploited group membership information to deanonymize social network users, showing that the groups a user joins are often sufficient to uniquely identify them.
More recently, \citet{ederer2024anonymity} deanonymized users of the Economics Job Market Rumors forum by exploiting weaknesses in its username generation scheme, recovering IP addresses for 66\% of posts.
These classical approaches required either structured features, exploitable technical vulnerabilities, or graph structure---none could operate on unstructured text at scale.

\paragraph{LLMs enable inference of personal attributes from unstructured text.}
The Extract step of our framework in \Cref{ssec:method_framework} follows a line of work of LLM inferring demographic data from online profiles.
\citet{staab2024beyond} show that LLMs can infer personal attributes such as location, occupation, and income from text with high accuracy, demonstrating privacy risks beyond training data memorization.
\citet{du2025automated} extend this line of work with AutoProfiler, a system of four specialized LLM agents that collaboratively extract sensitive attributes from pseudonymous platform activity.
They report 85-92\% accuracy in attribute extraction, demonstrating that automated profiling can be deployed at web scale.
However, their evaluation of actual deanonymization (linking extracted attributes to real identities) relies on manual LinkedIn searches and reports k-anonymity metrics rather than the high-precision matching we focus on.
Conversely, \citet{Staab2025LanguageMA} show that LLMs can be used to anonymize text while preserving utility, suggesting a potential defensive application of the same capabilities.
\edit{}{On the other hand, \citet{xin2025false} and \citet{krvco2026rat} show that even after removing explicit identifiers, contextual details in text still leak sensitive attributes, challenging the assumption that surface-level sanitization provides meaningful privacy.}
\edit{\citet{du2025beyond} survey emerging privacy risks from LLM deployment, distinguishing training-time data leakage from deployment-time threats such as automated profiling and social engineering.}{}

\edit{}{More broadly, the privacy community has begun shifting focus from training-data-centric concerns (memorization, extraction) toward deployment-time threats such as automated profiling and inference \citep{du2025beyond, mireshghallah2025position}.}
\edit{}{Our work provides large-scale empirical evidence for these deployment-time risks in the deanonymization setting.}

\paragraph{Recent work has begun exploring LLM-based deanonymization directly.}
\citet{li2026agentic} showed that agentic LLMs can deanonymize interview participants via web search, demonstrating that LLM agents make re-identification attacks low-effort: with a few natural-language prompts, off-the-shelf tools can search the web, cross-reference details, and propose matches.
Their work on the Anthropic Interviewer dataset~\citep{handa2025interviewer,anthropic2025interviewer} recovered 6 of 24 scientist identities by matching project descriptions to published papers, using task decomposition~\citep{jones2024adversaries} to bypass safeguards.
\edit{}{Apart from the mentioned AutoProfiler work on online platforms~\citep{du2025automated}, several other recent works discuss re-identification in the context of redacted datasets.}
\citet{nyffenegger2024anonymity} evaluate LLM re-identification capabilities on court decisions, finding that despite high re-identification rates on Wikipedia, even the best LLMs struggled with anonymized legal documents.
\edit{}{\citet{xin2025false} propose a framework for evaluating re-identification risk in sanitized text datasets (medical records and chatbot conversations), finding that commercial PII removal leaves enough semantic signal for an adversary with auxiliary information to infer sensitive attributes.}
Our work extends this line of research by developing a systematic framework for LLM-based deanonymization and evaluating it across multiple platforms and attack settings.

\paragraph{Stylometry identifies users through writing style rather than semantic content.}
Authorship attribution uses features such as function word distributions and syntactic structures~\citep{stamatatos2009survey}; recent work applies LLM embeddings to this task~\citep{tyo2022state}.
\edit{}{Determining whether two texts were written by the same person has been studied extensively through the PAN shared task series~\citep{stamatatos2023pan}, which evaluates pairwise verification on corpora of around 100 authors using stylometric features. Authorship attribution approaches explicitly treat topical similarity as a confound to be controlled for, whereas our method exploits semantic content as the primary signal. Our temporal-split Reddit setting (\cref{sec:study_reddit}) is related to authorship linking, but operates at larger scale (10{,}000 users) and relies on \emph{what} users say rather than just \emph{how} they say it.}
\edit{Our approach relies on semantic information---what users write about---rather than how they write; combining}{Combining stylometric and semantic approaches is an interesting direction for} future work.

\paragraph{More broadly, deanonymization is one of many ways LLMs empower both criminals and state actors.}
\citet{carlini2025llmsmonetize} argue that LLMs alter the economics of cyberattacks by enabling adversaries to launch tailored attacks on a user-by-user basis, fundamentally changing the cost-benefit calculus for attackers.
\citet{heiding2024spearphishing} demonstrate that LLM agents can autonomously crawl public information to construct profiles that were comprehensive for 88\% of targets, using them to generate spear phishing emails with click-through rates on par with human experts.

\section{Discussion}
\label{sec:discussion}

\paragraph{The future of online privacy.}
Deanonymization is one instance of LLMs acting as an ``information microscope'' that makes previously manual and expensive attacks scalable~\citep{hammond2025leviathan}.
Our paper shows that LLMs democratize deanonymization\edit{; the}{. Echoing concerns raised by prior work on LLM-based attribute inference and semantic privacy leaks~\citep{staab2024beyond, xin2025false, mireshghallah2025position}, we argue that the} asymmetry between attack cost and defense cost may force a fundamental reassessment of what can be considered private online. \edit{}{Our large-scale experiments provide quantitative evidence for these concerns in the deanonymization setting.}

So what do our findings mean for the future of privacy?
Governments could link pseudonymous accounts to real identities for surveillance of dissidents, journalists, or activists.
Corporations could connect seemingly anonymous forum posts to customer profiles for hyper-targeted advertising.
Attackers could build sophisticated profiles of targets at scale to launch highly personalized social engineering scams.
Hostile groups could identify important employees and decision makers and build online rapport with them to eventually leverage in various forms. Users, platforms, and policymakers must recognize that the privacy assumptions underlying much of today's internet no longer hold.

\paragraph{Possible mitigations.}
Enforcing a rate limit for API access to user data, detecting automated scraping, and restricting bulk data exports may reduce the severity of these attacks.
LLM providers could also monitor the use of their models to detect misuse such as deanonymization attempts~\citep{sumers2025protecting, anthropic2025disrupting}. Improved safety guardrails that make models refuse deanonymization requests might also provide some benefit, although our methods use LLMs in many ways that resemble benign usage (summarization of profiles, computing embeddings, etc.).

Classical anonymization frameworks such as $k$-anonymity~\citep{sweeney2002k} and differential privacy~\citep{dwork2006differential,dwork2014algorithmic} were designed for structured databases with explicit identifiers and assume attackers use direct matching or statistical queries. These frameworks do not account for the types of attacks we demonstrate; data releases should consider such threats when evaluating privacy risks. \edit{New frameworks could be developed with the threat of LLMs in mind.}{LLM-based text anonymization~\citep{Staab2025LanguageMA} offers a more targeted defense for unstructured text, though even these methods leave residual semantic signals~\citep{xin2025false}. 
}

\paragraph{Generalizability of our results.}
Our evaluation relies on ground truth datasets that may overestimate real-world success rates.
Users who publicly link their accounts or share identifying information may still share more information than they would for truly pseudonymous accounts, even considering our anonymization step.
However, measuring deanonymization performance requires ground truth, and we cannot verify matches for users who have not revealed their identities.
\edit{}{
While there are synthetic benchmarks for attribute inference \citep{yukhymenko2024synthetic}, 
it is not clear how to construct realistic synthetic datasets for deanonymization.
}
Profile splitting has different limitations: users might behave differently across truly separate platforms than across different communities or across time on the same site.
However, our methods work across a broad range of experimental setups, which suggests that they do generalize beyond any single evaluation setting.

Platforms should therefore assume that pseudonymous users can be linked across accounts and to real identities at scale. This should influence the decisions they make on data access policies. Users should similarly not assume that posting under a pseudonym provides meaningful protection.

\paragraph{Are our evaluations contaminated?}
One might wonder whether LLMs succeed at deanonymization because they memorized Reddit or Hacker News data during training. The fact that increasing reasoning effort substantially improves performance (\cref{sec:study_hn,sec:study_movies}) provides some evidence that reasoning plays a significant role and memorization alone cannot explain the results.
The training data for LLMs is typically not openly revealed, making it challenging to isolate these factors. We suspect that Hacker News and Reddit are part of most training corpora, but LinkedIn profiles are not. 
More broadly, even if memorization plays a role, this does not diminish the privacy implications: many social media platforms are included in LLM training corpora, so the deanonymization threat would only be reduced for platforms excluded from training data.

\section{Conclusion}

We have demonstrated that LLMs enable deanonymization of pseudonymous online accounts at scale, outperforming classical methods.
In many cases, LLMs enable us to perform attacks that would not have previously been possible, due to the lack of structured data or features.

\paragraph{These attacks require only publicly available models and standard APIs.}
Our pipeline uses only publicly available embedding models, standard LLM APIs, and LLM-agent scaffolding, placing them within reach of moderately resourced adversaries.

\paragraph{Pseudonymity does not provide meaningful protection online.}
Users who post under persistent usernames should assume that adversaries can link their accounts to real identities or to each other, and that the probability rises with each piece of micro-data they post.

\paragraph{Preventing such attacks appears challenging.} Not revealing any data on online platforms is difficult, as the data we use is the very data that makes online communities worthwhile. 
Although LLM providers could aim to detect and block attempts to misuse their models for deanonymization (as they do, for instance, for cyberattacks), we are pessimistic that this is possible as our deanonymization framework splits an attack into a combination of seemingly benign summarization, search and ranking tasks.

Recent advances in LLM capabilities have made it clear that there is an urgent need to rethink various aspects of computer security in the wake of LLM-driven offensive cyber capabilities. Our work shows that the same is likely true for privacy as well.

\section*{Ethical Considerations}

\paragraph{Stakeholder analysis.}
The primary stakeholders are social media users whose privacy could be compromised by deanonymization attacks.
General social media users who post under pseudonyms expecting privacy, vulnerable populations who depend on anonymity (activists, abuse survivors, whistleblowers), and the specific users whose data we used in experiments (Reddit users, HN users, etc.).
Secondary stakeholders include platforms, researchers, potential malicious actors, and society broadly.

\paragraph{Potential harms.}
The primary potential harm is that publishing this research could inspire malicious deanonymization attacks.
Concrete harms include stalking and harassment, doxxing of activists or vulnerable individuals, corporate surveillance and targeted manipulation, government surveillance and suppression of dissent, and chilling effects on free speech if people fear being identified.
However, these capabilities already exist in deployed LLMs; we are not introducing novel attack vectors but documenting existing risks.

\paragraph{Potential benefits.}
The primary benefit is raising awareness of privacy risks that already exist due to widely available LLM capabilities. 
Users can make informed decisions about what they share online and add better privacy measures.
Platforms can develop better privacy protections and reconsider making data publicly available, such as for LLM training. 
Policymakers can consider appropriate regulations, and LLM providers can consider adding additional safety guardrails that prevent large scale misuse.
The security community can develop defenses and metrics, similar to concepts such as k-anonymity. Before a dataset is irreversibly publicly released, researchers could study whether the information could be used by LLM agents to identify individuals.

\paragraph{Mitigations.}
We designed our experimental methodology to avoid directly harming individuals.
Most experiments do not deanonymize individuals, and we instead used synthetically constructed datasets (profile splitting, LLM-anonymized data). In the case of the Anthropic interview dataset, we note that a previous paper had already performed a similar attack~\citep{anthropic2025interviewer}.
We do not reveal any names or identities in this paper.
We do not release our matching pipeline code or processed datasets.

\paragraph{Decision to publish.}
We believe that the benefits of publication outweigh the marginal risks because these capabilities are already widely available.
Any moderately sophisticated actor can already do what we do using readily available LLMs and embedding models.
With future LLMs, without mitigations, this attack will be within the means of basically all adversarial actors.
By documenting the threat while it is nontrivial to execute, we enable proactive responses.
The privacy community, LLM providers, online platforms and users need to know about these risks, so not publishing would leave users unaware and unprotected.

\section*{Open Science}

Balancing the requirements of reproducibility while avoiding undue harm is challenging for works on deanonymization. Since defenses typically do not exist once the data is available, prior work has often refrained from releasing code to reproduce their attacks (e.g.,~\cite{narayanan2008deanonymization, cohen2022attacks, li2026agentic}).

Although many of our experiments are conducted with publicly available data that we synthetically anonymized for experimental purposes, we still believe that releasing this data and associated code would do more harm than is warranted by strict reproducibility concerns.
Indeed, the data consist of real user profiles that, while publicly available at the time of writing, should not be made available as research artifacts. Releasing code to directly run our attacks would needlessly make it trivial for someone to attempt a similar attack on other truly anonymous online profiles. 

\subsection*{Ethics}

This study was approved by ETH Zurich's Ethics Review Board (ERB).

\subsection*{Author Contributions and Acknowledgments}

This research was directed by ETH Zurich, which conducted the primary experiments. Nicholas Carlini advised on the research.
We thank Robin Staab and Mark Verö for extensive feedback.

\bibliography{main}
\bibliographystyle{plainnat}

\appendix
\crefalias{section}{appendix}
\crefalias{subsection}{appendix}

\section{Profile anonymization procedure}
\label{app:anonymization}

To prevent trivial deanonymization via unique identifiers while preserving semantic content relevant to our matching task, we apply anonymization rules based on whether searching the identifier online would directly reveal the profile owner.

\begin{table}[h]
\centering
\scriptsize
\begin{tabular}{@{}llll@{}}
\toprule
\textbf{Category} & \textbf{Example} & \textbf{Search Result} & \textbf{Action} \\
\midrule
Personal URL & knuth.stanford.edu & Finds Donald Knuth & Remove \\
Blog URL & karpathy.bearblog.dev & Finds Andrej Karpathy & Remove \\
Social handle & u/spez & Finds Steve Huffman & Remove \\
GitHub repo & flask & Finds Armin Ronacher & Remove \\
GitHub handle & torvalds & Finds Linus Torvalds & Remove \\
\midrule
Named bootcamp & ``F*****'' & Finds creator & Generalize \\
Founded project & ``C****'' & Finds founder & Generalize \\
\midrule
Mentioned colleague & Yann LeCun & Finds Meta & Keep \\
Local business & Cafe Carpe Diem & Finds the city & Keep \\
Institution & UCLA & Too broad & Keep \\
Demographics & Male, 40s & Too broad & Keep \\
Interests & plays chess & Too broad & Keep \\
Tech stack & uses LaTeX & Too broad & Keep \\
\bottomrule
\end{tabular}
\caption{Anonymization rules for profile data, closely matching our actual implementation. We remove direct identifiers, generalize unique project names, and keep information that does not uniquely identify the profile owner.}
\label{tab:anonymization}
\end{table}

\paragraph{Remove entirely.} We delete lines containing personal website/blog URLs, social media handles, and GitHub repository names or handles---any identifier where searching online directly reveals the profile owner.

\paragraph{Generalize.} We replace unique project names with generic descriptions (e.g., ``Founder of <specific programming community>'' becomes ``Founded a programming community'').

\paragraph{Keep as-is.} We retain colleague names (searching finds them, not the profile owner), local business names (searching finds the business, not its patrons), institution names, locations, demographics, interests, and technical skills.

\section{Agentic deanonymization experiments on Reddit}
\label{app:reddit_deanon}

We report full results for the two Reddit experiments summarized in \cref{sec:study_agents}.

\paragraph{Reddit academics.}
We curate 25 identifiable Reddit users posting on academic subreddits whose usernames reveal their real names. We manually verified each ground-truth identity, excluding ambiguous cases.
Then we anonymize their profiles to remove any direct identifiers, such as names or paper titles, but preserve any research interests or institutions they mention on their social media profile.
We run our de-anonymizing agent on this dataset, and it identifies 13 of 25 (52\%) with 72\% precision (5 incorrect, 7 abstentions).

\paragraph{Reddit career discussions.}
We collect 36 users who had posted their own LinkedIn URL for resume review on a software engineering career subreddit. We manually check each ground-truth identity, and exclude ambiguous cases.
After anonymizing their Reddit profiles, the agent correctly identifies 9 of 36 (25\%) at 90\% precision (1 incorrect, 26 abstentions).
The lower recall likely reflects that career discussion posts are less uniquely identifying: when we remove direct identifiers, many user's comments discuss generic software engineering topics without mentioning employers, projects, or credentials that would distinguish them from other users.

\section{Baseline attack details}
\label{app:narayanan_baseline}

We adapt the approach of \citet{narayanan2008deanonymization} to our datasets.

\paragraph{LinkedIn to Hacker News.}
For professional profile matching, each user is represented as a binary vector $\mathbf{x} \in \{0, 1\}^d$ where $d = 332$ is the vocabulary size, with each dimension corresponding to an attribute (e.g., \code{skill\_python}, \code{company\_google}). Given a query user with feature set $F_a$ and a candidate with feature set $F_b$, we compute weighted Jaccard similarity:
\[
\text{sim}(a, b) = \frac{\sum_{f \in F_a \cap F_b} w_f}{\sum_{f \in F_a \cup F_b} w_f}
\]
where the rarity weight $w_f = 1/\log(1 + c_f)$ downweights common attributes, and $c_f$ is the frequency of feature $f$ across all users. For each query user, we rank all candidates by descending similarity. For LinkedIn profiles, we parse the binary vector $\mathbf{x}$ directly from the structured data we collected; for HN profiles, we use an LLM with function calling to extract the same attributes from unstructured text.

\paragraph{Reddit movies.}
For movie review matching, we use a weighted similarity that accounts for rating agreement and temporal proximity:
\[
\text{sim}(a, b) = \frac{\sum_{m \in M_a \cap M_b} w_m \cdot s_m}{\sum_{m \in M_a \cup M_b} w_m}
\]
where $M_a, M_b$ are the sets of movies rated by users $a$ and $b$. The per-movie similarity combines a rating kernel and a temporal kernel:
\[
s_m = k_r^{\beta} \cdot k_t^{1-\beta}
\]
\[
k_r = \exp\left(-\frac{|r_a^m - r_b^m|}{\sigma_r}\right), \quad k_t = \exp\left(-\frac{|t_a^m - t_b^m|}{\sigma_t}\right)
\]
where $r_a^m$ is user $a$'s rating for movie $m$ and $t_a^m$ is the timestamp (in days). The rarity weight $w_m = 1/\log(1 + c_m)$ downweights popular movies, where $c_m$ is the total rating count for movie $m$. We use $\sigma_r = 1.0$, $\sigma_t = 40.0$ days, and $\beta = 0.5$.

Unlike \citet{narayanan2008deanonymization} who used a binary kernel (outputting 1 if reviews fall within a time threshold and have similar ratings, 0 otherwise), we use a smooth kernel because ratings and dates are heavily inferred by the LLM. Users rarely give explicit numerical ratings in Reddit comments, so the model must estimate scores from sentiment. Similarly, while we observe comment timestamps, a user's discussion of a single movie may span multiple comments over weeks or months, requiring the model to assign an approximate review date. Due to this inherent noise in LLM-extracted features, we use a more forgiving kernel that still rewards temporal and rating proximity.

\paragraph{Reddit temporal.}
For temporal profile matching, each user is represented as a binary vector $\mathbf{x} \in \{0,1\}^d$ where $d$ is the number of subreddits and $x_s = 1$ if the user posted in subreddit $s$.
Given a query user with subreddit set $S_q$ and a candidate with subreddit set $S_c$, we compute a weighted intersection score:
\[
\text{sim}(q, c) = \sum_{s \in S_q \cap S_c} w_s
\]
where the rarity weight $w_s = 1/\log |U_s|$ and $|U_s|$ is the number of users who posted in subreddit $s$.
This follows the original \nsattack formulation: rare subreddits (with fewer users) provide stronger evidence of a match. Unlike normalized metrics such as Jaccard or cosine similarity, this score is unbounded and favors users who post in many rare subreddits.

\section{LinkedIn to HN matching: additional details}
\label{app:linkedin_hn_additional}

\subsection{Scaling analysis}
\label{app:scaling}

We evaluate how recall at 90\% precision scales with candidate pool size by subsampling from our full 89k HN user pool. \Cref{tab:scaling} shows recall for Search and Reason (high) at different pool sizes.

\begin{table}[h]
\centering
\small
\caption{Recall @ 90\% precision by candidate pool size. Starred rows show log-linear extrapolated values.}
\label{tab:scaling}
\begin{tabular}{rcc}
\toprule
Candidate pool & Reason (high) & Search (embeddings) \\
\midrule
10 & 90.0\% & 80.0\% \\
100 & 82.0\% & 74.0\% \\
1k & 68.3\% & 47.6\% \\
10k & 63.2\% & 40.3\% \\
30k & 59.0\% & 35.1\% \\
89k & 55.2\% & 26.6\% \\
\midrule
1M* & \textit{45.1\%} & \textit{12.1\%} \\
10M* & \textit{36.2\%} & \textit{0.0\%} \\
100M* & \textit{27.4\%} & \textit{0.0\%} \\
\bottomrule
\end{tabular}
\end{table}

Both methods degrade log-linearly with pool size. We fit a log-linear model:
\[
\text{Recall}(\%) = a \cdot \log_{10}(N) + b
\]
where $N$ is the candidate pool size, fitted via ordinary least squares on the measured data points. For Reason: $a = -8.88$, $b = 98.35$; for Search: $a = -13.94$, $b = 95.73$. Extrapolated values for 1M, 10M, and 100M candidates are shown in italics in \cref{tab:scaling}. Reason degrades much more gracefully: at 100M candidates it would still retain an estimated 27\% recall at 90\% precision, while Search falls to zero.

\section{Reddit movies: additional details}
\label{app:reddit_movies_additional}

\subsection{Recall by shared movies}
\label{app:recall_by_shared_movies}

\Cref{tab:recall_by_shared_movies} provides a detailed breakdown of recall at different precision thresholds, stratified by the number of movies a user discussed in both r/movies and the alternative movie communities. Users with more shared movies are substantially easier to identify.

\begin{table}[htbp]
\centering
\small
\caption{Recall at precision thresholds by number of shared movies (total queries: 9,781). 95\% Wilson CIs shown.}
\label{tab:recall_by_shared_movies}
\begin{tabular}{@{} lrcc @{}}
\toprule
\#Shared & n & 90\% Prec & 99\% Prec \\
\midrule
1 & 4,729 & 3.1\% & 1.2\% \\
 & & \footnotesize (2.6--3.6) & \footnotesize (0.9--1.6) \\[0.3em]
2--4 & 3,693 & 8.4\% & 2.5\% \\
 & & \footnotesize (7.6--9.4) & \footnotesize (2.0--3.0) \\[0.3em]
5--9 & 1,118 & 23.2\% & 7.1\% \\
 & & \footnotesize (20.8--25.7) & \footnotesize (5.7--8.7) \\[0.3em]
10+ & 241 & 48.1\% & 17.0\% \\
 & & \footnotesize (41.9--54.4) & \footnotesize (12.8--22.3) \\[0.3em]
Overall & 9,781 & 8.5\% & 2.8\% \\
 & & \footnotesize (8.0--9.1) & \footnotesize (2.4--3.1) \\
\bottomrule
\end{tabular}
\end{table}

\subsection{Reasoning effort comparison}
\label{app:reasoning_comparison}

\Cref{tab:recall_at_precision} compares recall at different precision thresholds when using low versus high reasoning effort on GPT-5.2. High reasoning effort consistently outperforms low reasoning, roughly doubling recall at 99\% precision (2.8\% vs.\ 1.4\%).

\begin{table}[h]
\centering
\small
\caption{Recall at different precision thresholds for Reddit movies matching (9,781 queries) by reasoning effort. Reason uses GPT-5.2. 95\% Wilson CIs shown.}
\label{tab:recall_at_precision}
\begin{tabular}{@{} lccc @{}}
\toprule
Method & 90\% Prec & 98\% Prec & 99\% Prec \\
\midrule
Reason (low) & 5.2\% & 2.0\% & 1.4\% \\
 & \footnotesize (4.8--5.7) & \footnotesize (1.7--2.3) & \footnotesize (1.1--1.6) \\[0.3em]
Reason (high) & \textbf{8.5\%} & \textbf{3.8\%} & \textbf{2.8\%} \\
 & \footnotesize (8.0--9.1) & \footnotesize (3.4--4.2) & \footnotesize (2.4--3.1) \\
\bottomrule
\end{tabular}
\end{table}

\section{Comparing calibration methods: confidence scores vs.\ sorting}
\label{app:tournament_movies}
\label{app:calibration_comparison}

There are two ways to use LLMs in the Calibrate step of our framework: (1) ask the LLM to output confidence scores directly alongside its match decision, or (2) sort all proposed matches by running pairwise comparisons.
We use the first approach in \cref{sec:study_hn,sec:study_movies}, where the LLM \edit{outputs a confidence score via function calling}{outputs a structured confidence score} during the Reason step, and we threshold this score to trace precision-recall curves.
We use the second approach---Swiss-system sorting with Bradley-Terry ratings---in \cref{sec:study_reddit}, where it yields large gains over embedding similarity as a confidence measure.

These two approaches differ in what information they exploit.
Direct confidence scoring evaluates each query-candidate pair independently: the LLM sees a single proposed match and estimates how plausible it is.
Pairwise sorting compares proposed matches against each other: the LLM sees two query-candidate pairs side by side and judges which is more plausible.
Sorting thus has access to cross-query information that direct scoring does not, but it also requires a large batch of queries and incurs additional LLM calls (approximately $O(n \log n)$ comparisons for $n$ proposed matches).

\paragraph{Movie matching experiment.}
We attempted to apply pairwise sorting to the Reddit movie matching task from \cref{sec:study_movies}. On top of all matches confirmed by the high-reasoning GPT-5.2 verification stage, we ran 15 rounds of pairwise comparisons using \texttt{gpt-5-mini-2025-08-07}.

Results were worse than using the verification-stage confidence scores directly. At 99\% precision, sorting achieved only 0.7\% recall compared to 3.1\% for confidence ranking. At 90\% precision, sorting achieved 3.6\% recall versus 8.7\% for confidence. At 80\% precision, sorting achieved 7.0\% recall versus 10.5\% for confidence.

We note that the verification stage uses a much more powerful model (GPT-5.2 with high reasoning effort) than sorting (GPT-5-mini). Additionally, sorting's $O(n \log n)$ LLM calls make it less suitable for expensive models or larger dataset sizes. These results suggest that when the Reason step already uses a strong model with high reasoning effort, sorting provides limited additional benefit.

\section{Linking temporally split Reddit profiles: additional details}

\subsection{Data processing pipeline}
\label{app:reddit_pipeline}

We start from publicly available Reddit comment dumps from Academic Torrents, then apply a multi-stage filtering pipeline to select suitable authors and comments. We first scan all comment dumps to aggregate per-author statistics (total comment count, date range of activity), then apply the following filters:
\begin{enumerate}[nosep]
    \item Activity span: at least 1{,}095 days (3 years) between first and last comment.
    \item Minimum volume: at least 200 total comments.
    \item Rate limit: at most 24 comments per day on average. We find that this simple heuristic is highly effective in excluding automated accounts and bots.
    \item Bot exclusion: usernames ending in \code{bot}, \code{gpt}, or \code{mods} are removed, along with a manually curated bot list.
\end{enumerate}
This yields an initial filtered author set. We then extract all comments for these authors and build an index of comment timestamps per author.

We then split each author's comment history into two disjoint profiles separated by a temporal gap.
For each author, we find the split time $t^*$ with gap window $g = 365$ days which maximizes the number of comments in the smaller of the resulting halves, subject to a minimum of 100 comments on each side. Comments within the gap, i.e., in the interval $[t^* - g/2,\; t^* + g/2]$, are discarded.
This produces two temporally separated split-profiles per author:
\begin{itemize}[nosep]
    \item Query half (before): all comments with timestamp $< t^* - g/2$.
    \item Candidate half (after): all comments with timestamp $> t^* + g/2$.
\end{itemize}
The gap ensures that the matching task is more difficult and more realistic. We exclude authors who do not satisfy the 100-comment minimum on both sides, and shuffle the remaining authors with a fixed random seed for reproducible sampling. We select 15{,}000 authors for our primary dataset.

Comments are then deduplicated and filtered for relevance before feature extraction. During data loading, duplicate comments (same author and timestamp) are dropped. We apply a two-stage relevance filter when running the LLM-based embedding method:
\begin{enumerate}[nosep]
    \item \textbf{Heuristic pre-filter}: removes deleted/removed comments, ultra-short comments ($\leq$3 characters), and pure URL comments.
    \item \textbf{LLM-based semantic filter}: batches of comments are sent to a smaller LLM (Gemini 3 Flash) that identifies generic or non-personal content (e.g., one-word replies in large default subreddits). Each comment is labeled as relevant or irrelevant. Only relevant comments are retained for summarization.
\end{enumerate}

\subsection{Prompts}
\label{app:prompts}

\paragraph{Filtering.}
The following prompt is used with Gemini 3 Flash to pre-filter generic and uninformative comments.

\begin{tcolorbox}[colback=gray!5, colframe=gray!50, boxrule=0.5pt, title={\small Filtering Prompt}, breakable]
  \ttfamily\small
  Your task is to identify Reddit comments that should be filtered out. All comments below are from the SAME author, listed in CHRONOLOGICAL ORDER (oldest first). A comment should be FILTERED if:\\
1. It reveals nothing about the author, OR\\
2. It reveals information already shown in an earlier comment (redundant)\\

CRITICAL: Check the subreddit FIRST\\

Before looking at comment content, check if the subreddit itself reveals information. If YES, KEEP the FIRST comment from that subreddit, but FILTER subsequent comments from the same revealing subreddit (redundant - we already know they participate there).\\

REVEALING SUBREDDITS (KEEP first occurrence, FILTER subsequent):\\
- Health/medical: r/diabetes, r/celiac, r/adhd, r/anxiety, r/depression, etc.\\
- Cities/regions: r/nyc, r/seattle, r/Portland, r/austin, r/london, r/toronto, etc.\\
- Professions: r/teachers, r/nursing, r/medicine, r/law, r/sysadmin, etc.\\
- Religion: r/atheism, r/christianity, r/islam, r/judaism, etc.\\
- Identity: r/actuallesbians, r/gaybros, r/lgbt, r/transgender, r/childfree, etc.\\
- Sports teams: r/bostonceltics, r/lakers, r/49ers, r/cowboys, etc.\\
- Niche hobbies: r/woodworking, r/knitting, r/homebrewing, r/beekeeping, etc.\\

GENERAL-INTEREST SUBREDDITS (subreddit alone reveals nothing):\\
- r/AskReddit, r/funny, r/pics, r/videos, r/gaming, r/movies, r/music, r/aww, etc.\\

Then check comment content for NEW information\\

FILTER if:\\
- Generic reactions: "lol", "this", "same", "agreed", "nice", "wow", "thanks"\\
- Information already revealed in an earlier comment (e.g., second mention of "my wife")\\

KEEP if it reveals NEW information about:\\
- Demographics (age, gender, location, occupation)\\
- Personal circumstances ("my wife", "my kids", "when I was in college")\\
- Opinions or values (political, religious, lifestyle stances)\\
- Expertise or interests shown through knowledge\\

EXAMPLE (same author):\\
{[0]}  [r/Portland] Just moved here last year\\{}
[1] [r/diabetes] My A1C is finally under control\\{}
[2] [r/Portland] The food carts are amazing\\{}
[3] [r/gaming] I'm 34 and still play regularly\\{}
[4] [r/diabetes] Same here, diet really helped\\{}
[5] [r/pics] lol\\{}
[6] [r/AskReddit] My wife and I love hiking\\{}
[7] [r/videos] My wife showed me this\\{}

Output: 2,4,5,7\\

Explanation:\\
- [0] KEEP: r/Portland reveals location (first occurrence)\\{}
- [1] KEEP: r/diabetes reveals health condition (first occurrence)\\{}
- [2] FILTER: r/Portland redundant (already know they're in Portland from [0])\\{}
- [3] KEEP: reveals age\_30s (new info)\\{}
- [4] FILTER: r/diabetes redundant (already know from [1])\\{}
- [5] FILTER: generic "lol" in general-interest sub\\{}
- [6] KEEP: reveals married (new info)\\{}
- [7] FILTER: "my wife" redundant (already know from [6])\\{}

Output ONLY the indices of comments to FILTER. Comma-separated, no spaces (e.g. 0,3,7). If none should be filtered, output "none".

COMMENTS:
{comments}
\end{tcolorbox}

\paragraph{Summarization.}
The following prompt is used with Gemini 3 Pro to summarize split profiles based on their comments.

\begin{tcolorbox}[colback=gray!5, colframe=gray!50, boxrule=0.5pt, title={\small Summarization Prompt}, breakable]
  \ttfamily\small
Below are comments from a single Reddit user. Output a comma-separated list of DISTINCTIVE traits that would help identify this specific person among thousands of random Reddit users.\\

CRITICAL: Focus on traits that are BOTH distinctive AND persistent over time. If you saw this user's comments from a different year, these traits should still be true.\\

SKIP these (too common to be useful):\\
- uses\_profanity, analytical\_tone, sarcastic\_tone, informal\_tone, verbose\_style\\
- Generic "gaming\_interest" (but DO include specific games like "dwarf\_fortress\_player")\\
- Country-level locations like "lives\_in\_usa" (but DO include city/region like "lives\_in\_portland")\\
- One-time mentions or temporary states (job\_hunting, currently\_sick, just\_moved)\\

PRIORITIZE distinctive persistent traits:\\
- Specific locations (city/region level): lives\_in\_portland, lives\_in\_nelson\_bc\\
- Profession field: aerospace\_engineer, pediatric\_nurse, high\_school\_teacher\\
- Core identity: gender, has\_children, homeowner, has\_dog, has\_cat\\
- Deep hobbies (not one-off mentions): woodworking, homebrewing, keeps\_reptiles\\
- Strong values: pro\_nuclear, vegan, libertarian, atheist, gun\_owner\\
- Specific fandoms: dwarf\_fortress\_player, formula1\_fan, liverpool\_fc\_supporter\\
- Life circumstances: expat, military\_veteran, recovering\_alcoholic\\

Format:\\
- Lowercase, 1-4 words, underscores between words\\
- Be thorough - more traits is better than fewer
- Order by distinctiveness (most identifying first)\\

Examples of GOOD output:\\
lives\_in\_portland, male, aerospace\_engineer, has\_dog, woodworking\_interest, pro\_nuclear, recovering\_alcoholic, nfl\_fan, atheist\\

lives\_in\_uk, female, nurse, has\_children, marathon\_runner, plays\_guitar, labour\_supporter\\

Examples of BAD output (too generic or unstable):\\
male, age\_30s, lives\_in\_usa, married, employed, gaming\_interest, uses\_profanity, analytical\_tone, job\_hunting\\

Now output the descriptors.
\end{tcolorbox}

The following is a synthetic example summary illustrating the output format. It does not correspond to any real user.

\begin{tcolorbox}[colback=gray!5, colframe=gray!50, boxrule=0.5pt, title={\small Example Output (Synthetic)}]
\ttfamily\small
lives\_in\_nelson\_bc, pediatric\_nurse, keeps\_ball\_pythons, homebrews\_mead, female, married, has\_two\_daughters, roller\_derby\_player, prius\_owner, sourdough\_obsessed, plays\_stardew\_valley, critical\_role\_fan, ndp\_voter, pro\_nuclear, celiac\_disease, plays\_mandolin, thru\_hiked\_pct, volunteer\_firefighter, dislikes\_cilantro
\end{tcolorbox}

\paragraph{Record selection.}
The following prompt is used with Gemini 3 Pro to perform LLM-based record selection using the top-k candidates.

\begin{tcolorbox}[colback=gray!5, colframe=gray!50, boxrule=0.5pt, title={\small Record Selection Prompt}, breakable]
  \ttfamily\small
You are matching Reddit users across two time periods. Given a QUERY user profile and {k} CANDIDATE profiles, determine which candidate (if any) is the same person as the query.\\

QUERY profile:\\
{query\_summary}\\

CANDIDATE profiles:\\
{candidates}\\

Which candidate is the same person as the query? Consider overlapping traits like location, profession, hobbies, demographics, and values. A match should share multiple distinctive traits, not just one or two common ones.\\

Output ONLY a single number (1-{k}) for the best match.
\end{tcolorbox}

\paragraph{Sorting (pairwise comparisons).}
The following prompt is used with \texttt{gpt-5-mini-2025-08-07} to perform pairwise comparisons between query-candidate pairs for confidence sorting.

\begin{tcolorbox}[colback=gray!5, colframe=gray!50, boxrule=0.5pt, title={\small Pairwise Comparison Prompt}, breakable]
\ttfamily\small
\# Task\\
You are helping to match Reddit accounts. We have two accounts: 'query' and 'candidate'. The goal is to determine if a query-account and candidate-account belong to the same person.\\

You will see two matches (MATCH A and MATCH B). Each match proposes that a specific query-account and candidate-account are the same person. Your job is to decide which match is more plausible.\\

\# Data format\\
Each account is summarized as a comma-separated list of tags describing the user (e.g., location, occupation, interests, writing style).\\

<few-shot examples>\\

\#\# MATCH A\\
Query: <candidate\_a query summary>\\
Candidate: <candidate\_a target summary>\\

\#\# MATCH B
Query: <candidate\_b query summary>\\
Candidate: <candidate\_b target summary>\\

\# Response options
- **A**: MATCH A is the better match
- **B**: MATCH B is the better match

Respond with: A, B
\end{tcolorbox}

\subsection{Recall@Precision as a function of the fraction of matchable queries $\matchfrac$}
\label{app:precision_pi}

Let $M$ be the number of matchable queries, $N$ the number of non-matchable queries, and $\matchfrac = \frac{M}{M + N}$.
Define $\text{TP}$ as the number of true positives (correct attacker guesses), $\text{FP}_M$ as the number of false positives for matchable queries (wrong attacker guess when there exists a true match), and $\text{FP}_N$ as the number of false positives for non-matchable queries (the attacker does not abstain for users without a true match). With this notation, the three key rates are
\begin{itemize}
    \item True Positive Rate/Recall: $\tpr = \frac{\text{TP}}{M}$
    \item False Match Rate: $\fmr = \frac{\text{FP}_M}{M}$
    \item False Positive Identification Rate: $\fpir = \frac{\text{FP}_N}{N}$ .
\end{itemize}

Precision can be rewritten in terms of those rates and $\matchfrac$ as follows:
\begin{align*}
    \precision &= \frac{\text{TP}}{\text{TP} + \text{FP}_M + \text{FP}_N} \\
    &= \frac{
        \matchfrac \frac{\text{TP}}{M}
    }{
        \matchfrac \frac{\text{TP}}{M}
        + \matchfrac \frac{\text{FP}_M}{M}
        + (1- \matchfrac) \frac{\text{FP}_N}{N}
    } \\
    &= \frac{\matchfrac \cdot \tpr}{\matchfrac \cdot \tpr + \matchfrac \cdot \fmr + (1-\matchfrac) \cdot \fpir} \, .
\end{align*}
Since all three rates depend on \emph{either} the $M$ matchable queries or the $N$ non-matchable queries, but \emph{not both},
we can estimate all rates independently of $\matchfrac$.

\subsection{Justifying Reasoning and Calibration for temporally split Reddit profiles}
\label{app:method_justification}

\begin{figure*}[t]
    \centering
    \begin{subfigure}[t]{\figsixcol}
        \centering
        \includegraphics[width=\textwidth]{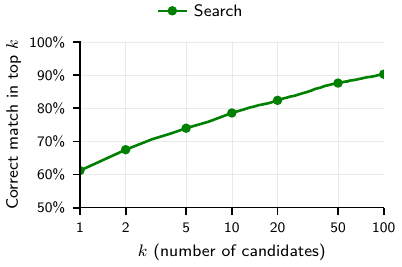}
        \caption{Top-k recall for Search}
        \label{fig:topk_recall}
    \end{subfigure}
    \begin{subfigure}[t]{\figsixcol}
        \centering
        \includegraphics[width=\textwidth]{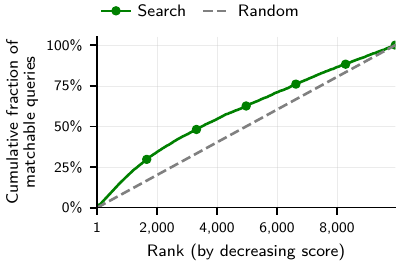}
        \caption{Fraction of matches containing a matchable query when ranked by similarity score}
        \label{fig:a_fraction_comparison}
    \end{subfigure}
    \caption{Similarity score is good at narrowing down the candidate set (a) but not at ranking matches in terms of plausibility (b).}
    \label{fig:method_justification}
\end{figure*}

\paragraph{Justifying Reasoning.}
For every k, we plot the fraction of queries where the top-k candidates in terms of embedding similarity contain the true match
in \cref{fig:topk_recall}.
While rank-1 accuracy is only around 60\%, this rises to over 80\% at $k = 15$.
Thus, embedding similarities effectively narrow down the candidate pool to a small size (such that using LLMs with reasoning for candidate selection becomes feasible), but fail to identify the one true match consistently.

\paragraph{Justifying Calibration.}
In \cref{fig:a_fraction_comparison}, we take the maximum query-candidate embedding similarity for all query users (5k with a true match, 5k with no true match), sort queries in decreasing similarity, and show the fraction of matchable users at every rank.
The resulting curve barely surpasses random ordering.
Thus, embedding similarity is poorly calibrated, justifying the use of LLMs for better calibration.

\subsection{Results}
\label{app:results}

\begin{table}[h]            
  \centering
  \small                                                                                         
  \caption{Recall at precision thresholds for Reddit split-profile matching (5k queries, 10k
  candidates).}                                                                                  
  \label{tab:reddit_recall_at_precision}                                                       
  \begin{tabular}{lccc}
  \toprule
  Method & 90\% Prec & 98\% Prec & 99\% Prec \\
  \midrule
  Baseline (Narayanan) & 0.4\% & 0.2\% & 0.2\% \\[0.3em]
  LLM & 38.3\% & 20.2\% & 16.0\% \\[0.3em]
  LLM + Selection & 47.8\% & 28.2\% & 20.3\% \\[0.3em]
  LLM + Selection + Sorting & 67.3\% & 47.6\% & 38.4\% \\
  \bottomrule
  \end{tabular}
  \end{table}

\begin{figure*}[t]
    \centering
    \begin{subfigure}[t]{\figsixcol}
        \centering
        \includegraphics[width=\textwidth]{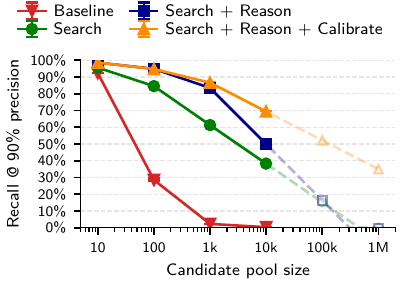}
        \caption{Effects of candidate set size at 90\% precision}
        \label{fig:recall_vs_candidate_size_p90}
    \end{subfigure}
    \hfill
    \begin{subfigure}[t]{\figsixcol}
        \centering
        \includegraphics[width=\textwidth]{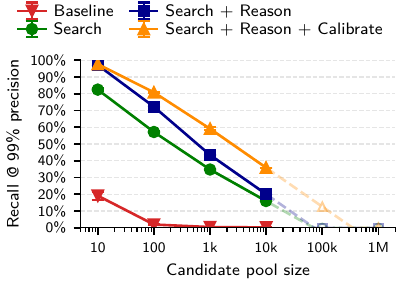}
        \caption{Effects of candidate set size at 99\% precision}
        \label{fig:recall_vs_candidate_size_p99}
    \end{subfigure}
    \caption{Full results for \cref{fig:reddit_ablations_candidates}.}
\end{figure*}

\begin{figure*}[t]
    \centering
    \begin{subfigure}[t]{\figsixcol}
        \centering
        \includegraphics[width=\textwidth]{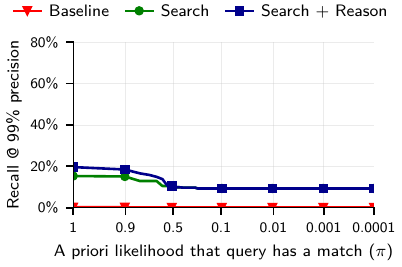}
        \caption{Effects of the likelihood that a query has a match ($\matchfrac$) at 99\% precision}
        \label{fig:recall_vs_pi_p98}
    \end{subfigure}
    \caption{Full results for \cref{fig:reddit_ablations_matchfrac}.}
\end{figure*}
\end{document}